\documentclass[aps,prd,twocolumn,amsmath,amssymb,floatfix,nofootinbib,superscriptaddress]{revtex4-1}

\usepackage{graphicx}
\usepackage{dcolumn}
\usepackage{bm}
\usepackage[usenames, dvipsnames]{color}
\usepackage{mhchem}
\usepackage[normalem]{ulem}
\graphicspath{ {figures/} }
\usepackage{epstopdf}
\usepackage[usenames]{color}
\usepackage[colorlinks,bookmarks]{hyperref}
\definecolor{linkblue}{rgb}{0,0,0.8}
\definecolor{linkgreen}{rgb}{0,0.5,0}
\definecolor{purple}{rgb}{0.78,0.18,0.77}

\hypersetup{pdfpagemode=UseNone, pdfstartview=FitH, linkcolor=linkblue, %
            citecolor=linkgreen, urlcolor=linkblue}

\newcommand{\be}{\begin{eqnarray}}
\newcommand{\non}{\nonumber \\}
\newcommand{\ee}{\end{eqnarray}}

\def\beq{\begin{equation}}
\def\eeq{\end{equation}}
\newcommand{\lp}{\left(}
\newcommand{\rp}{\right)}
\newcommand{\lb}{\left[}
\newcommand{\rb}{\right]}

\newcommand{\msun}{M_\odot}
\newcommand{\NHD}{N_{\rm{HD}}}
\newcommand{\xHD}{x_{\rm{HD}/\rm{H}2}}
\newcommand{\Px}{P_{\rm{HD}\times\rm{[CII]}}}

\usepackage[export]{adjustbox}

\newcommand{\apjl}{Astrophys. J. Lett.}
\newcommand{\aap}{Astron. Astrophys.}
\newcommand{\apjs}{Astrophys. J. Suppl. Ser.}

\newcommand{\mnras}{Mon. Not. R. Astron. Soc.}
\newcommand{\jcap}{J. Cosm. Astropart. Phys.}
\newcommand{\pasp}{Publ. Astron. Soc. Pac.}

\begin{document}

\title{Mapping the Universe in HD}
\newcommand{\cita}{Canadian Institute for Theoretical Astrophysics, University of Toronto, 60 St.~George Street, Toronto, M5S 3H8, Canada}
\newcommand{\crc}{Canada Research Chair in Theoretical Astrophysics}
\newcommand{\nyu}{Center for Cosmology and Particle Physics, Department of Physics, New York University, 726 Broadway, New York, NY, 10003, U.S.A.}
\newcommand{\perimeter}{Perimeter Institute for Theoretical Physics, 31 Caroline Street North, Waterloo, ON N2L 2Y5, Canada}
\newcommand{\drao}{Dominion Radio Astrophysical Observatory, Herzberg Astronomy \& Astrophysics Research Centre, National Research Council Canada, P.O.~Box 248, Penticton, BC V2A 6J9, Canada}
\newcommand{\smu}{Department of Physics,
Southern Methodist University, 3215 Daniel Ave, Dallas, TX 75275, U.S.A.}
\newcommand{\aip}{Leibniz-Institut f\"ur Astrophysik Potsdam (AIP), An der Sternwarte 16, D-14482 Potsdam, Germany}

\author{Patrick~C.~Breysse}
\affiliation{\nyu}
\author{Simon Foreman}
\affiliation{\perimeter}
\affiliation{\drao}
\author{Laura~C.~Keating}
\affiliation{\aip}
\author{Joel Meyers}
\affiliation{\smu}
\author{Norman Murray}
\affiliation{\cita}
\affiliation{\crc}

\date{\today}

\begin{abstract}

Hydrogen deuteride (\ce{HD}) is prevalent in a wide variety of astrophysical environments, and measuring its large-scale distribution at different epochs can in principle provide information about the properties of these environments. In this paper, we explore the prospects for accessing this distribution using line intensity mapping of emission from the lowest rotational transition in HD, focusing on observations of the epoch of reionization ($z\sim6-10$) and earlier. We find the signal from the epoch of reionization to be strongest most promising, through cross-correlations within existing [CII] intensity mapping surveys.  While the signal we predict is out of reach for current-generation projects, planned future improvements should be able to detect reionization-era HD without any additional observations, and would help to constrain the properties of the star-forming galaxies thought to play a key role in reionization.
We also investigate several avenues for measuring HD during ``cosmic dawn" ($z\sim10-30$), a period in which HD could provide one of the only complementary observables to 21$\,$cm intensity maps. 
We conclude that existing and planned facilities are poorly matched to the specifications desirable for a significant detection, though such a measurement may be achievable with sustained future effort. Finally, we explain why HD intensity mapping of the intergalactic medium during the cosmic dark ages ($z\gtrsim 30$) appears to be out of reach of any conceivable experiment.

\end{abstract}

\maketitle


\section{Introduction}

Line-intensity mapping, the measurement of the integrated flux of spectral lines emitted from galaxies or the intergalactic medium, is a rapidly growing field with wide-ranging applications in cosmology and astrophysics~\cite{Kovetz:2017agg,Kovetz:2019uss}.  By observing spatial fluctuations in the emission, intensity-mapping surveys are able to make use of the emission of many unresolved galaxies, rather than identifying individual objects above some flux cut. Accurate redshifts can typically be obtained due to the sharp spectral features which are observed, thereby allowing surveys to produce three-dimensional maps of structure of the Universe as traced by the total flux of some emission line.

Several spectral lines have been identified as potential targets of line-intensity mapping surveys~\cite{Visbal:2011ee}. The 21-cm line resulting from the spin flip transition of neutral hydrogen has been the focus of a great deal of study (see \cite{Pritchard:2012im,Ansari:2018ury} for reviews), while the \ce{CO} rotational lines, the [CII] fine-structure line, and the Ly$\alpha$ line have begun to receive increased attention~\cite{Kovetz:2017agg,Kovetz:2019uss}.

Molecular hydrogen (\ce{H2}) is by far the most abundant molecule in the Universe.  Molecular gas is the main driver of star formation, and knowledge of its density is key to understanding the interstellar medium, star formation history, and galaxy evolution~\cite{Kennicutt:2012ea}.  Unfortunately, \ce{H2} possesses no permanent electric dipole moment and therefore has only quadrupolar rotational transitions ($\Delta J=2$).  The lowest of these rotational states lies about 510~K above ground, and the spontaneous decay time is on the order of 100~yr, making it a faint line visible only in hot environments.  These factors make \ce{H2} a challenging target for line-intensity mapping, and we will quantify just how challenging as part of this work. 

Carbon and oxygen are among the most abundant elements in galaxies, and combine to form \ce{CO} in molecular clouds. \ce{CO} possesses a modest dipole moment (0.11~D) and low excitation energy for the ground rotational transition ($h\nu / k_{\rm B} = 5.53$~K), which means that even in the cold environments of molecular clouds, \ce{CO} rotational states are easily excited.  The emission from rotational transitions of \ce{CO} also falls in convenient atmospheric windows (115.27 GHz or 2.6 mm for the 1-0 transition), making \ce{CO} an excellent candidate for line-intensity mapping surveys.  The prevalence and observability of \ce{CO} have made it a common observational target, and have motivated many attempts to use the \ce{CO} density as a tracer of the molecular gas density through the \ce{CO}-to-\ce{H2} conversion factor~\cite{Bolatto:2013co}.  This conversion factor has a relatively large systematic uncertainty in our own galaxy, depends on the environment, and is less certain in other galaxies and at higher redshifts.  \ce{CO} intensity mapping surveys have the potential to probe the properties of high-redshift galaxies on the faint end of the luminosity function and to study the gas density and star formation rate of these poorly understood objects~\cite{Li:2015gqa,Breysse:2015saa}. 

Here we discuss another potential line-intensity mapping target which is abundant, prevalent, and directly traces the density of molecular hydrogen.  Hydrogen deuteride (\ce{HD}) has a weak permanent electric dipole moment~\cite{Wick:1935hd} due to the proton-deuteron mass difference, which causes the electrons to orbit more closely around the latter, giving the ground state $8.56\times 10^{-4}$~D~\cite{Pachucki:2008hdd}.  The 1-0 rotational transition of \ce{HD} lies at 2.675 THz (112 $\mu$m), which is strongly absorbed in the atmosphere, making it very difficult to observe from the ground.  The excitation energy for the ground rotational state is fairly high ($h\nu / k_{\rm B} = 128$~K) compared to that of $\ce{CO}$.  Rotational transitions of \ce{HD} in emission have been observed with ISO~\cite{Wright:1999iso,Bertoldi:1999iso,Polehampton:2002iso}, UKIRT~\cite{Howat:2002uki}, the Spitzer Space Telescope~\cite{Neufeld:2006spi,Yuan:2012spi}, and the Herschel Space Observatory~\cite{Bergin:2013lka}, and have also been seen in absorption spectra of quasars~\cite{Ivanchik:2015xia} and within the Milky Way (e.g.~\cite{Spitzer:1973,Spitzer:1974,Morton:1975,Lacour:2004dg,Snow:2008st}). The HD(1-0) line is also a key target of the Origins Space Telescope for measuring the masses of protoplanetary disks \cite{Battersby2018}.

Deuterium was produced in the first few minutes of the radiation-dominated era in the process of Big Bang Nucleosynthesis (BBN)~\cite{Cyburt:2015bbn}.  There are no known astrophysical sources of deuterium, and so essentially all deuterium in the Universe is of primordial origin~\cite{Epstein:1976hq,Prodanovic:2003bn}.  Deuterium is burned in stars, and so the deuterium abundance has decreased monotonically from its primordial value throughout the history of the Universe.  The primordial deuterium abundance has been measured to be $(\ce{D}/\ce{H})_p = (2.527 \pm 0.030) \times 10^{-5}$~\cite{Cooke:2017cwo} in good agreement with the value predicted theoretically by standard BBN~\cite{Pitrou:2018cgg} when using the photon-to-baryon ratio as determined from measurements of the cosmic microwave background (CMB) made by the Planck satellite~\cite{Ade:2015xua}.

Hydrogen deuteride formed well before reionization, and significant fractionation resulted in a freeze-out ratio of $\ce{HD}/\ce{H2} \approx 7 \times 10^{-4}$ by $z \approx 40$, a factor of about 25 larger than the primordial $\ce{D}/\ce{H}$ ratio~\cite{Galli:2012rf}.  \ce{HD} is therefore expected to be present with a significant abundance in essentially all galaxies at all redshifts, even in the pristine primordial environments of the first collapsed objects (depending on environmental factors, such as the ionizing background radiation at the relevant epoch).

Several upcoming surveys aim to perform intensity mapping of the [CII] fine structure line at 1.901~THz (158~$\mu$m) over a wide range of redshifts (see~\cite{Kovetz:2017agg} for a summary).  These same experiments are capable of doing intensity mapping with the \ce{HD}(1-0) rotational line in a higher redshift window.  For example, the Fred Young Submillimeter Telescope (FYST,~\cite{Herter2019}\footnote{\url{www.ccatobservatory.org}}) is capable of observing the redshifted [CII] line in the window $5<z<9$, which would allow for intensity mapping of \ce{HD} in the range $7.5<z<13$.  It is particularly useful that there is a range of redshifts $7.5<z<9$ where both [CII] and \ce{HD} will be visible with FYST, which allows for cross correlations using maps obtained by the same instrument.

In this work, we investigate the prospects for detection of an HD(1-0) line intensity signal from three different eras: reionization ($z\sim6-10$), cosmic dawn ($z\sim10-30$), and the dark ages ($z\gtrsim30$). A measurement from each era would provide distinct information: the signal from reionization would tell us about the star-forming galaxies thought to play a key role in the creation of ionizing bubbles of radiation; the signal from cosmic dawn would tell us about the conditions in the minihalos that host the first generation of stars; and the signal from the dark ages would provide us with information about density fluctuations in the intergalactic medium, which would directly trace cosmological perturbations in the same way envisioned for dark-ages 21$\,$cm measurements~\cite{Loeb:2003ya}.

We find the signal from reionization to be the most promising for detection: while likely out of reach of the planned [CII] survey by FYST, a more ambitious ``Stage II" [CII] survey would be capable of ${\rm SNR} \sim 11$ for even our most pessimistic model. While this number corresponds to raw statistical significance only, ignoring the details of foreground subtraction and other systematics, we nevertheless see it as an encouraging sign that \ce{HD}(1-0) may be observable by intensity mapping surveys intended for [CII].

For cosmic dawn, we examine several observational strategies, and conclude that existing and planned facilities are poorly matched to the specifications desirable for making an intensity map of \ce{HD}(1-0): only several years' worth of ALMA observing time over a 0.01 deg$^2$ patch or CMB-HD~\cite{Sehgal:2019ewc} observing time over a few-deg$^2$ patch would have any chance of detecting the \ce{HD}(1-0) auto power spectrum. Possibilities for cross-correlations are limited, and will not fare any better with current or planned experiments. However, a futuristic spectroscopic CMB satellite, such as the enhanced-PIXIE configuration envisioned for observing time evolution of the CMB blackbody temperature~\cite{Abitbol:2019ewx}, would likely have the statistical power for an auto spectrum measurement, assuming appropriate control over practicalities such as foreground separation.

Finally, \ce{HD}(1-0) intensity mapping in the dark ages appears to be completely infeasible, due primarily to the high energy (compared to 21$\,$cm in \ce{H}I) required to excite \ce{HD} into the first rotational state, along with the low abundance of \ce{HD} (again compared to \ce{H}I).

This paper is organized as follows. In Sec.~\ref{sec:reion}, we present a model for the \ce{HD}(1-0) line intensity auto spectrum, and cross spectrum with [CII], based on radiative transfer and estimates of the total abundance of molecular gas, and forecast their detectability by FYST and a next-generation [CII] intensity mapping survey. In Sec.~\ref{sec:cd}, we adapt the halo model for \ce{H2} intensity mapping from Ref.~\cite{Gong:2012iz} to \ce{HD} at cosmic dawn, and discuss a variety of possible observing strategies. In Sec.~\ref{sec:darkages}, we estimate the strength of \ce{HD}(1-0) emission from the intergalactic medium during the dark ages, and then we  conclude in Sec.~\ref{sec:conclusion}. Appendices~\ref{app:MofM} and~\ref{app:cdforecasts} contain extra details about our forecasts, while Appendix~\ref{app:h2} consists of \ce{H2} cosmic dawn intensity mapping forecasts, based on Ref.~\cite{Gong:2012iz} but with updated model ingredients and instrumental specifications.

For all computations, we use cosmological parameters from the Planck 2015 results, given in the ``TT,TE,EE+lowP+lensing+ext" column of Table~4 of Ref.~\cite{Ade:2015xua}.

\section{Reionization}
\label{sec:reion}

\subsection{Radiative Transfer}
Here we present an estimate of the HD intensity mapping signal during reionization, based on simple radiative transfer arguments. 

Consider a line of sight through an HD-emitting galaxy with total optical depth $\tau$ in a given rotational transition.  A cosmologically-nearby observer will see an intensity
\be
I_{\rm{HD}}=\int_0^\tau \frac{j_{\rm{HD}}(\tau')}{\kappa_{\rm{HD}}(\tau')}e^{-\tau'}d\tau',
\label{eq:radtransfer}
\ee
where $j_{\rm{HD}}$ and $\kappa_{\rm{HD}}$ are the emission and absorption coefficients and we have neglected background radiation and assumed that our observation has a frequency resolution wider than the target line.  For a line of width $\Delta\nu$ in local thermodynamic equilibrium (LTE), $j_{\rm{HD}}/\kappa_{\rm{HD}}=B_\nu(T)\Delta\nu$, where $B_\nu$ is the Planck function at excitation temperature $T$.  We then have
\beq
I_{\rm{HD}}=B_\nu(T)\Delta\nu\left(1-e^{\tau}\right)=B_\nu(T)\Delta\nu\left(1-e^{N_\ell\sigma_{\rm{HD}}}\right),
\label{eq:IHD1}
\eeq
 where $N_{\ell}$ is the column density of HD molecules in the lower state of our transition and $\sigma_{\rm{HD}}$ is the cross-section for the transition, given by~\cite{Spitzer1978}
\be
\sigma_{\rm{HD}}=\frac{3c^2A_{u\ell}}{8\pi\nu_{\rm{HD}}^2\Delta\nu}\left(1-e^{-h\nu_{\rm{HD}}/k_{\rm B}T}\right).
\ee
The Einstein coefficient $A_{u\ell}$ for the HD(1-0) transition is $5.1\times10^{-8}\ \rm{s}^{-1}$~\cite{Flower2000}.
We assume the line width $\Delta\nu$ is dominated by turbulent motions, and adopt a value corresponding to a velocity width of $\Delta v=10$ km/s, comparable to local giant molecular clouds \citep{Solomon1987}.

Under the continued assumption of LTE, the total column density $N_{\rm{HD}}$ of HD molecules can be inferred from
\be
\frac{N_\ell}{\NHD}=\frac{2J+1}{Z_{\rm{HD}}}e^{hB_{\rm{HD}}J(J+1)/k_{\rm B}T},
\ee
for a transition with lower rotational quantum number~$J$.  The rotational constant $B_{\ce{HD}}$ is $1.33\times10^{12}\ \rm{s}^{-1}$~\citep{Endres2016}.
The partition function $Z_{\ce{HD}}$ takes its usual form
\be
Z_{\ce{HD}} = \sum_{J=0}^\infty(2J+1)e^{-hB_{\rm{HD}}J(J+1)/k_{\rm B}T}.
\label{eq:partitionfun}
\ee
Combining Equations (\ref{eq:IHD1}-\ref{eq:partitionfun}) yields
\be
\nonumber
I_{\rm{HD}}&=&\frac{2h\nu_{\rm{HD}}^4\Delta v}{c^3}\frac{1-\exp(\NHD\sigma_{\rm{HD}}/Z_{\rm{HD}})}{\exp(h\nu_{\rm{HD}}/k_{\rm B}T)-1} \\
&\approx&C_{NI}(T)\NHD,
\ee
where we have defined
\be
C_{NI}(T)\equiv\frac{2h\nu_{\rm{HD}}^4\sigma_{\rm{HD}}(T)\Delta v}{c^3Z_{\rm{HD}}(T)\left[\exp(h\nu_{\rm{HD}}/k_{\rm B}T)-1\right]},
\ee
and assumed that the HD transition is optically thin in all regimes of interest.

We can verify this latter assumption by considering a typical molecular column through the Milky Way.  If we define the HD/H$_2$ ratio $\xHD\equiv \NHD/N_{\text{H2}}$, we can express the HD optical depth in terms of the surface density $\Sigma_{\rm{H2}}$ of molecular hydrogen.  In the extreme case wherein all of the HD is in the lower rotational state (i.e. $N_\ell=N_{\rm{HD}}$), the optical depth will be
\be
\tau_{\rm{HD}}(N_\ell=N_{\rm{HD}})=\xHD\sigma_{\rm{HD}}\frac{\Sigma_{\rm{H2}}}{m_{\rm{H2}}},
\ee
where $m_{\text{H2}}$ is the mass of a single \ce{H2} molecule. We adopt $\xHD=10^{-4}$ for the purposes of this section, which is broadly consistent with the models from Ref.~\cite{McGreer:2008wf}.  For a typical Milky Way line of sight with $\Sigma_{\rm{H2}}\sim10$ $M_{\odot}$ pc$^{-2}$~\cite{2017ApJ...834...57M}, we obtain
\be
\tau_{\rm{HD}}(N_\ell=N_{\rm{HD}})\sim5\times10^{-4}\left(\frac{\Sigma_{\rm{H2}}}{10\ M_{\odot}/\rm{pc}^2}\right).
\ee
Because the transition is optically thin, the HD luminosity $L_{\rm{HD}}$ of a galaxy will simply scale as its total HD mass.  Thus, again assuming a constant $\xHD$, we can write
\be
L_{\rm{HD}}=4\pi C_{NI}(T)\xHD\frac{M_{\text{H2}}}{m_{\text{H2}}},
\label{eq:lhd}
\ee
for a galaxy with total H$_2$ mass $M_{\rm{H2}}$.

If we want to know the amplitude of the intensity mapping power spectrum, we need to integrate over the full distribution of halo luminosities,
\beq
\label{eq:barI}
\bar{I}_{\rm{HD}}(z) = \int_{M_{\rm min}}^{M_{\rm max}} dM \frac{dn}{dM}(z)
	\frac{L_{\rm{HD}}(M,z)}{4\pi D_{\rm L}(z)^2} y_{\ce{HD}}(z) D_{\rm A}(z)^2\ ,
\eeq
which, using Eq.~\eqref{eq:lhd}, corresponds to integrating over the molecular gas content of the halos. In Eq.~\eqref{eq:barI}, $M_{\rm min}$ and $M_{\rm max}$ are the minimum and maximum masses of halos emitting in the given line, $dn/dM$ is the halo mass function, $D_{\rm L}$ is the luminosity distance, $D_{\rm A}$ is the comoving angular diameter distance (equal to the comoving radial distance $\chi$), and
\beq
y_{\ce{HD}}(z)\equiv \frac{d\chi}{d\nu_{\ce{HD}}} = \frac{\lambda_{\ce{HD}} (1+z)^2}{H(z)}
\eeq
translates between $\chi$ and frequency, where $\lambda_{\ce{HD}}$ is the rest wavelength of \ce{HD}(1-0).
It is difficult to model the full distribution $M_{\text{H2}}(M)$, due to uncertainties about conditions in high-redshift galaxies. However, in the absence of this information, we can reduce Eqs.~(\ref{eq:lhd}-\ref{eq:barI}) to
\beq
\bar{I}_{\rm{HD}}(z)=\frac{C_{NI}(T)\xHD y_{\rm{HD}}(z) D_{\rm A}(z)^2\rho_c(z)}{m_{\rm{H}2}D_{\rm L}(z)^2}\Omega_{\rm{H}2}(z),
\label{eq:IHD}
\eeq
where $\rho_c$ is the critical cosmological density and $\Omega_{\rm{H}2}$ is the total fraction of that density contributed by molecular gas.  Thus we can estimate the strength of the HD signal from the total molecular gas content of the universe.

Several of the quantities that go into Eq.~\eqref{eq:IHD} are highly uncertain even in the local universe, and will be even more so at high redshift.  It is therefore beyond our capabilities to claim a single estimate for the HD intensity.  We will instead adopt a range of parameter values in an attempt to roughly quantify the range of possible signals.  We will focus on two parameters which have a large impact on $\bar{I}_{\rm{HD}}$: the cosmological molecular gas abundance $\Omega_{\rm{H2}}$ and the excitation temperature $T$.  

\begin{figure}
\centering
\includegraphics[width=\columnwidth]{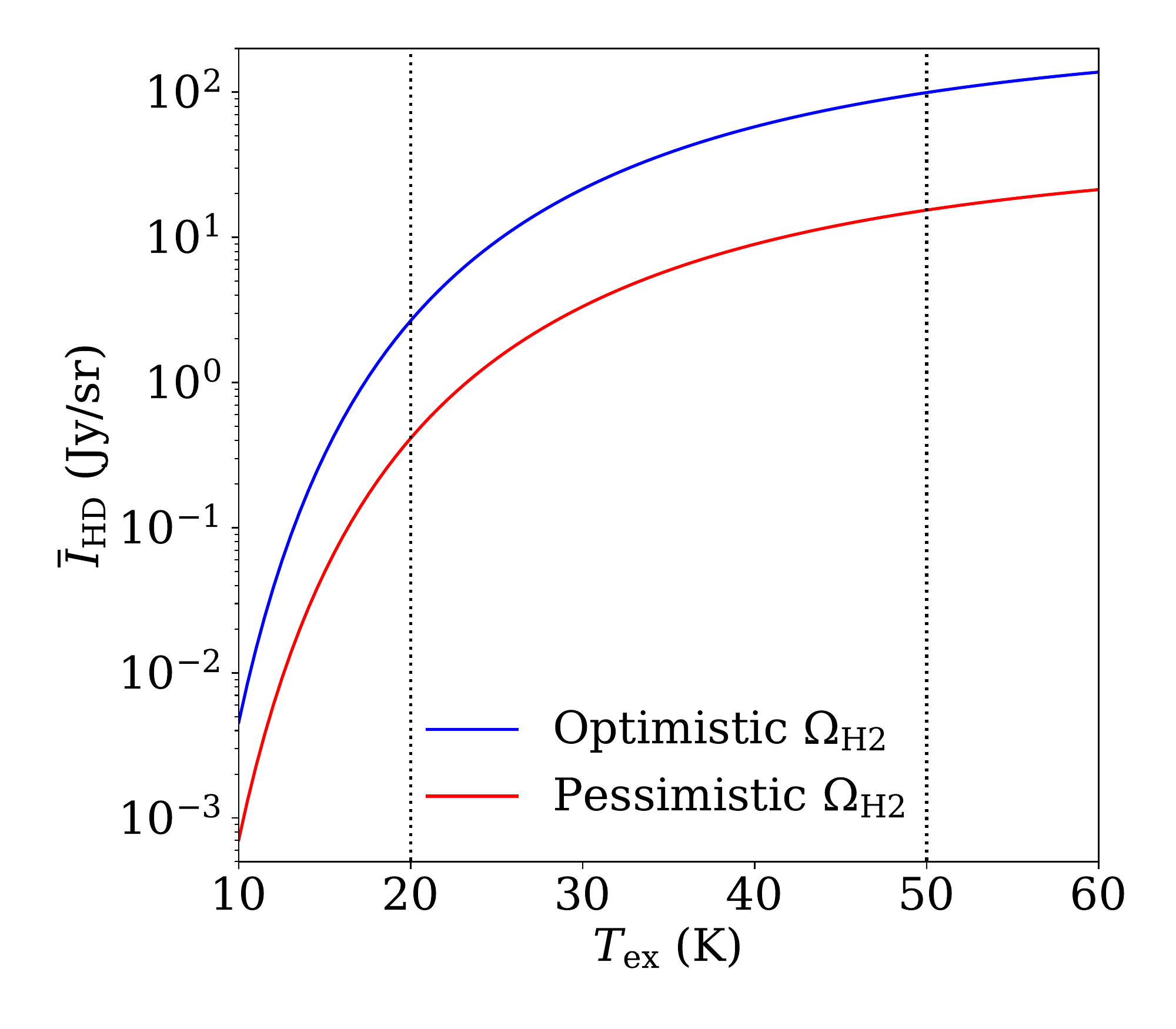}
\caption{Sky-averaged mean intensity of the HD line at $z=6$ computed using Eq. (\ref{eq:IHD}).  Results are shown as a function of excitation temperature, with the optimistic and pessimistic values we assume marked with dotted lines.  The blue and red curves assume the optimistic and pessimistic $\Omega_{\rm{H2}}$ values respectively from Ref. \cite{Lagos2011}.}
\label{fig:IvsT}
\end{figure}

For the molecular gas abundance, we rely on the simulated results presented in Ref.~\cite{Lagos2011}.  Their Fig.~17 presents a range of possible $\Omega_{\rm{H2}}$ values as a function of redshift.  For our optimistic and pessimistic models we will use the highest and lowest values from these simulations, which correspond to the ``Bau05.BR" and ``Bow06.KMT" values\footnote{The main differences between these models relate to how star formation is suppressed in massive galaxies, the assumed stellar mass function for starbursts, and the implementation of star formation; see Ref.~\cite{Lagos2011} for details.} (at $z\sim 6$, these are $\Omega_{\rm{H2}} \approx 7\times 10^{-4}$ and $1\times 10^{-4}$). 
 For the excitation temperature, we choose $T=20$~K for the pessimistic value and $T=50$~K for the optimistic model.  The lower value is broadly consistent with conditions in local molecular clouds, while the higher is in rough agreement with the dust temperature obtained from the Planck collaboration's modelling of the cosmic infrared background at $z\sim6$~\cite{Planck2014}.  Fig.~\ref{fig:IvsT} shows the dependence of $\bar{I}_{\rm{HD}}$ on these two parameters, illustrating the range of models we consider.  This clearly demonstrates the large amount of uncertainty on the strength of the HD signal, with $\sim2$ orders of magnitude between the brightest and faintest intensities.

\subsection{The Reionization-Era Power Spectrum}

Intensity maps are typically analyzed in terms of their power spectra.  The power spectrum of line intensity fluctuations is the sum of contributions from large-scale clustering and Poisson shot noise:
\begin{align}
\label{eq:pclus}
P_X^{\rm clus}(k,z) &= \bar{I}_X(z)^2 \bar{b}_X(z)^2 P_{\rm m}(k,z)\ , \\
\label{eq:pshot}
P_X^{\rm shot}(z) &= \int_{M_{\rm min}}^{M_{\rm max}} dM \frac{dn}{dM}(z) \non
&\qquad\times
	\lb \frac{L_X(M,z)}{4\pi D_{\rm L}(z)^2} y_X(z) D_{\rm A}(z)^2 \rb^2\ ,
\end{align}
where we have used $X$ to denote both the particle type and emission line we are considering (e.g.~\ce{HD}(1-0)).
The shape of the clustering term is set by the matter power spectrum $P_{\rm m}$, and is weighted by the mean line intensity and the luminosity-weighted bias
\beq
\bar{b}_X(z) = \frac{\int_{M_{\rm min}}^{M_{\rm max}} dM \frac{dn}{dM}(z) L_X(M,z) b(M,z)}
	{\int_{M_{\rm min}}^{M_{\rm max}} dM \frac{dn}{dM}(z) L_X(M,z)}\ .
\label{eq_bavg}
\eeq 
At the redshifts we consider here ($z\gtrsim5$ or so), we expect structure growth to be relatively linear on scales that are clustering-dominated rather than shot-noise-dominated.  We thus expect the one-halo contribution to the power spectrum to be subdominant to the linear and shot-noise components.  Further, any one-halo term will certainly be small compared to the uncertainty in our modelling.  Thus, we neglect this term in our forecasts. We take $M_{\rm{min}}=10^8\ M_{\odot}$ in this section, though as shown in Appendix \ref{app:MofM} our power spectra are relatively insensitive to this choice.  We choose an arbitrarily large value of $M_{\rm{max}}=10^{15}\ M_{\odot}$ that is larger than the most massive halos we expect to see at these redshifts. 

In order to fully model the power spectrum of HD fluctuations, we would need to understand how the global abundance of molecular gas is distributed in halos of different mass.  The shot noise power in an HD survey would be
\begin{multline}
P^{\rm{shot}}_{\rm{HD}}(z)=C_{NI}^2(T)\left(\frac{\xHD}{m_{\rm{H2}}}\right)^2\\
\times\int dM\frac{dn}{dM}(z) \left[\frac{M_{\rm{H2}}(M)}{D_L^2(z)}y_{\rm{HD}}(z)D_A^2(z)\right]^2,
\end{multline}
and the luminosity-weighted bias for HD would be given by
\be
\bar{b}_{\rm{HD}}=\frac{\int dM M_{\rm{H2}}(M) b(M) dn/dM}{\int dM M_{\rm{H2}}(M) dn/dM}.
\ee
Estimating the full behavior of these quantities would require a functional form for $M_{\rm{H2}}(M)$, which we do not obtain from our radiative transfer model.  However, given the uncertainty demonstrated above, the exact impact of changing $M_{\rm{H2}}(M)$ will likely be small compared to the difference in $\bar{I}_{\rm{HD}}$ values over the parameter ranges we consider (see Appendix~\ref{app:MofM}). We will thus make the simplifying assumption that $M_{\rm{H2}}(M)\propto M$ for the purposes of this section.  

The HD(1-0) transition during the epoch of reionization falls into the frequency range of several upcoming intensity mapping experiments seeking to map reionization with the 158 $\mu$m [CII] transition, including FYST, the Tomographic Intensity Mapping Experiment (TIME, \cite{Crites2014}), and the CarbON [CII] line in post-rEionization and ReionizaTiOn epoch project (CONCERTO, \cite{Lagache2018}). This coincidence means intensity maps of reionization-era HD will already exist in the data taken by these telescopes.  Even in our most optimistic models, the HD line will be subdominant to the target [CII] line in all of these surveys.  However, we can extract the HD line by constructing cross-correlations between different frequency channels in these maps.  As described in Ref.~\cite{Breysse2017}, the HD and [CII] lines will trace the same large-scale structure at different observed frequencies, so by correlating different frequency bands we can isolate both lines at the same redshift.  The observed cross-spectrum in this case takes the form
\begin{align}
\nonumber
\Px(k, z) &= \bar{I}_{\rm{HD}}(z) \bar{I}_{\rm{[CII]}}(z) 
\bar{b}_{\rm{HD}}(z) \bar{b}_{\rm{[CII]}}(z) P_m(k, z) \\
&\quad +P^{\rm{shot}}_{\rm{HD}\times\rm{[CII]}}(z),
\end{align}
where $\bar{I}_{\rm{[CII]}}$ and $\bar{b}_{\rm{[CII]}}$ are the mean intensity and bias of the [CII] line, computed analogously to those of HD.  The shot power in a cross-correlation between two intensity mapping lines is given by
\begin{align}
P_{\rm{HD}\times\rm{[CII]}}^{\rm shot}(k,z) &= \int_{M_{\rm min}}^{M_{\rm max}} dM \frac{dn}{dM}(z) L_{\rm{HD}}(M)L_{\rm{[CII]}}(M)\non
&\qquad\qquad\times
	\lb \frac{y_X(z) D_{\rm A}(z)^2}{4\pi D_{\rm L}(z)^2} \rb^2\ ,
\end{align}
as derived in Ref.~\cite{Liu2020}.  

The error on the cross-power spectrum between two intensity maps is
\begin{widetext}
\beq
\label{eq:sigmacross}
\sigma_{\times}(k)=\frac{1}{\sqrt{N_{\rm{modes}}(k)}}
\left[\Px^2(k)W_{\rm{HD}}(k)W_{\rm{[CII]}}(k) 
+\left(P_{\rm{HD}}(k)W_{\rm{HD}}(k)+P^N_{\rm{HD}}\right)
\left(P_{\rm{[CII]}}(k)W_{\rm{[CII]}}(k)+P^N_{\rm{[CII]}}\right)\right]^{1/2},
\eeq
\end{widetext}
where the $W_X(k)$ factors give the suppression of the signal at high $k$ due to the finite instrument resolution  and at low $k$ due to the finite survey area~\cite{Bernal2019}, and
\be
N_{\rm{modes}}(k)=\frac{k^2\Delta kV_{\rm{surv}}}{4\pi^2}
\ee
is the number of independent Fourier modes available in a bin of width $\Delta k$ for a survey covering total comoving volume $V_{\rm{surv}}$, accounting for the fact that only half of these modes are independent for a real field.
The noise power spectrum of a map of $X$, denoted by $P_X^N$ in Eq.~\eqref{eq:sigmacross}, is given by Eqs.~(\ref{eq:pN}-\ref{eq:Vvox}) in Appendix~\ref{app:h2} (we use $\Omega_{\rm pix}=\theta_{\rm FWHM}^2$ in Eq.~\ref{eq:tpix} for the surveys in this section). 

The total signal-to-noise ratio (SNR) obtained over all $k$ will then be
\beq
SNR=\left[\sum_k\frac{\Px^2(k)W_{\rm{HD}}(k)W_{\rm{[CII]}}(k)}{\sigma^2_\times(k)}\right]^{1/2}.
\label{eq:SNR}
\eeq
Note that Eqs.~\eqref{eq:sigmacross} and~\eqref{eq:SNR} only hold in the approximation of Gaussian statistics in both intensity maps at the scales of interest; this will not be true in the shot noise regime, but the beam suppression factors $W_X(k)$ imply that the information obtainable from this regime is subdominant from that at larger scales.

Of the currently planned or in-progress [CII] experiments described above, FYST has the largest total frequency coverage, and thus offers the widest overlap between the [CII] and HD lines, so we will use it to represent the constraining power of current experiments.  To demonstrate possibilities for future observations, we will also forecast for a lightly-modified version of the [CII] Stage II survey presented in Ref. \cite{Silva2015}.  The parameters we assume for these two surveys can be found in Table~\ref{tab:EoRspecs}.  In the FYST frequency range, we get the best overlap between the two lines for a cross-correlation centered at $z=6$.  For FYST, we use the most up-do-date projections for their Deep Spectroscopic Survey\footnote{Dongwoo Chung, private communication, see also Stacey et al. in prep.}.

For the [CII] Stage II survey, we adjust the target frequency range as well as the frequency and angular resolutions to match FYST for ease of comparison.  For both surveys, we model [CII] emission using model ``m1" from Ref.~\cite{Silva2015}, though it should be noted that the strength of the [CII] signal will likely be similarly uncertain.

\begin{table}
\centering
\begin{tabular}{l c c c c}
\hline
\ & 
FYST & 
FYST &
Stage II &
Stage II \\
\ & HD & [CII] & HD & [CII] \\
\hline
\hline\\[-0.5em]
$\sigma_{\rm{pix}}$ (MJy sr$^{-1/2}$ s$^{1/2}$) & 2.1 & 0.50 & 0.044 & 0.089 \\
$n_{\rm{pix}}$ & 20\footnote{FYST has $\sim6000$ total detectors, but the effective number is reduced due to their use of a Fabry-Perot interferometer which cannot observe every frequency channel simultaneously \cite{Chung:2018szp}.} & 20\textsuperscript{a} & 16000 & 16000 \\
$\theta_{\rm{FWHM}}$ (arcsec) & 35 & 50 & 35 & 50 \\
$\nu_{\rm{obs}}$ (GHz) & 380 & 267 & 380 & 267 \\
$\Delta \nu$ (GHz) & 100 & 71 & 100 & 71 \\
$\delta \nu$ (GHz) & 3.8 & 2.7 & 2.5 & 2.5 \\
$t_{\rm{obs}}$ (hr) & 4000 & 4000 & 4000 & 4000 \\
$\Omega_{\rm{surv}}$ (deg$^2$) & 8.0 & 8.0 & 2.0 & 2.0 \\
[+0.5em]
\hline
\end{tabular}
\caption{\label{tab:EoRspecs} Parameters for our reionization-era HD-[CII] cross-correlation forecasts at $z=6$, including the overall sensitivity $\sigma_{\rm{pix}}$, the number of spatial pixels $n_{\rm{pix}}$, the beam full width at half max $\theta_{\rm{FWHM}}$, the observing frequency $\nu_{\rm{obs}}$, the overall frequency bandwidth $\Delta \nu$, the channel width $\delta\nu$, the total observing time $t_{\rm{obs}}$, and the total survey area $\Omega_{\rm{surv}}$.}
\end{table}

\begin{figure}
\centering
\includegraphics[width=\columnwidth]{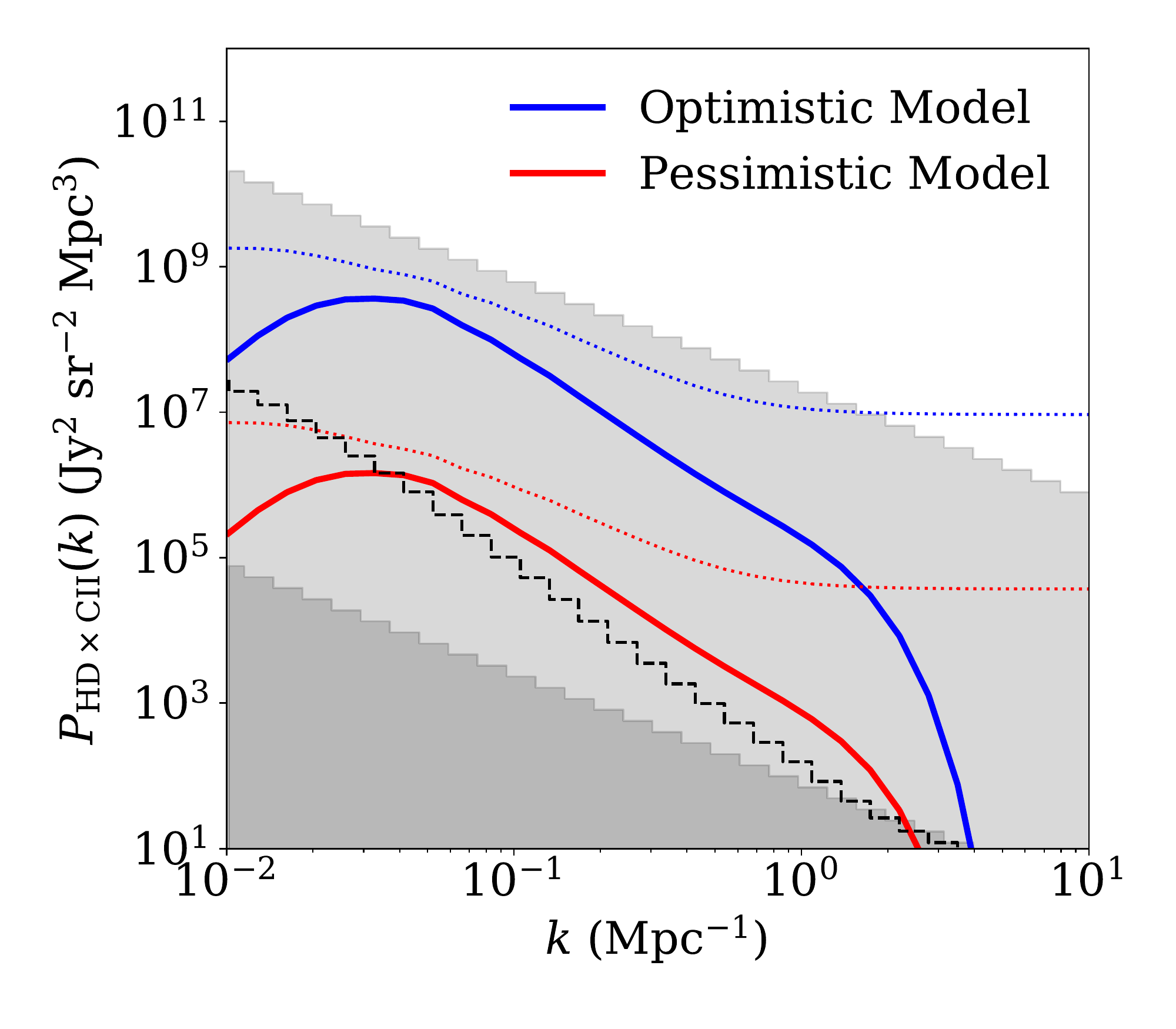}
\caption{[CII]-HD cross-power spectra at $z=6$ assuming [CII] model ``m1" from Ref. \cite{Silva2015} and the optimistic (blue) and pessimisic (red) HD models described above.  Dotted lines show the full, unsmoothed power spectra, solid lines show the spectra smoothed by the instrument resolution $W(k)$.  Shaded regions show the sensitivity limits of the FYST (light shading) and [CII] Stage II (dark shading) surveys defined in Table \ref{tab:EoRspecs}.  Noise levels assume Fourier bins of width $\Delta\log k=0.1$, and the noise curves are plotted as ``staircases" to show the widths of these bins. The gray bands show only the instrumental contribution to the power spectrum error, with no sample variance term included.  For the sake of comparison, the black dotted line shows the total (noise + sample variance) error on the pessimistic model assuming the [CII]-Stage II setup.
}
\label{fig:EoR_Pk}
\end{figure}

Fig.~\ref{fig:EoR_Pk} shows the power spectra obtained by our optimistic and pessimistic radiative transfer models, compared to the sensitivities of the two experiments.  For the FYST survey, we obtain an all-$k$ SNR of 0.3 for the brightest of the two models.  This puts the signal out of reach of current-generation experiments, though given the immense uncertainty in the signal amplitude even an upper limit may be interesting.  For future experiments, we find that even the fainter of the two models is detectable with a SNR of $\sim11$, making HD an excellent unique tracer of molecular gas during cosmic reionization. Our forecasts do not include systematics such as foreground subtraction, but we expect these systematics to affect [CII] and HD similarly, so that more detailed treatments for [CII] elsewhere in the literature should apply equally well to a [CII]-HD cross-correlation.

\section{Cosmic Dawn}
\label{sec:cd}

\subsection{Gas cooling by HD}

An intensity mapping signal in \ce{HD}(1-0) may also be generated by the halos that act as sites for formation of Population III stars. To see how this might arise, we first review the basic scenario for how these stars form (see Refs.~\cite{Glover:2012gx,Bromm:2013iya} for reviews).

The first stars are typically expected to form at $z\sim 20-30$ in dark matter halos with $M\sim 10^6\msun$, based on the criteria that (1) the gravitationally bound gas is able to cool efficiently enough to collapse down to a protostellar core, and (2) the cosmic density field has a sufficient number of peaks that can collapse into halos of the appropriate mass. The virial temperature of the gas in such halos, $T_{\rm vir} \sim 2\times 10^3\,  [(1+z)/20]\,{\rm K}$, is well below the threshold, $\sim10^4\,{\rm K}$, at which substantial cooling can take place via atomic transitions. However, the residual free electrons left over from recombination are sufficient to catalyze the formation of \ce{H2} with an abundance of $10^{-4}-10^{-3}$, enough to allow the gas to cool down to roughly $200\,{\rm K}$ at $n\sim 10^4\,{\rm cm}^{-3}$. Once this limit (set by the lowest allowed rotational transition in \ce{H2}, $510\,{\rm K}$) is reached, the rotational and vibrational levels in \ce{H2} attain their LTE populations, and the \ce{H2} cooling rate scales only linearly with number density (as opposed to the $n^2$ scaling at higher temperatures). From this point, cooling proceeds fairly slowly until the density reaches $10^8\,{\rm cm}^{-3}$, when three-body reactions begin to convert the rest of the \ce{H} into \ce{H2}, and the interplay between different heating and cooling processes causes the gas to collapse further.

This picture changes somewhat when the role of \ce{HD} is considered. The main chemical reaction that produces \ce{HD} in the gas is exothermic, whereas the inverse reaction is endothermic; this leads to fractionation, which boosts the \ce{HD}-to-\ce{H2} ratio above the primordial \ce{D}-\ce{H} ratio, with a stronger boost at lower temperatures. If \ce{H2} cools the gas to a sufficiently low temperature, enough \ce{HD} can be produced that it will become the dominant coolant. Due to the smaller energy required to excite its first rotational state ($128\,{\rm K}$), \ce{HD} can then allow the gas to cool almost down to the CMB temperature, until its critical density of $10^6\,{\rm cm}^{-3}$ is reached.

Several factors determine whether this will happen in a given halo. Simulations have shown that if the initial ionization fraction of the gas is no larger than that of the intergalactic medium (IGM) ($x_0\sim 2\times 10^{-4}$), halos with $M\gtrsim 10^6\msun$ will not form enough HD to significantly affect the gas cooling process~\cite{Bromm:2001bi,Nakamura:2002rj,Johnson:2005hk,Yoshida:2006bz,Ripamonti:2007hj,McGreer:2008wf,Kreckel:2010}, while HD-dominated cooling is seen in halos with the same ionization fraction but lower mass~\cite{McGreer:2008wf} or in halos that initially contain more ionized gas~\cite{Nagakura:2005ks,McGreer:2008wf,Johnson:2005hk,Greif:2008qqa,Hirano:2013lba,Nakauchi:2014pma,Hirano:2015wxa}. (The stars formed in these cases are often referred to as Population III.1, when the gas is of primordial composition, and Population III.2, when the gas has been partially ionized or otherwise affected by previous generations of objects.)

There are several possible mechanisms for increased ionization in these halos, including shocking by mergers or supernovae~\cite{Johnson:2005hk,Shchekinov:2005pz,Nakauchi:2014pma,Greif:2008qqa,Prieto:2013hpa}, proximity to relic HII regions~\cite{Johnson:2005hk,Nagakura:2005ks,Yoshida:2006dv}, or the influence of far-ultraviolet background radiation or cosmic rays~\cite{Jasche:2007gq,Nakauchi:2014pma,Hirano:2015wxa}. On the other hand, a UV background may instead act to dissociate \ce{HD} and/or \ce{H2}, preventing molecular cooling from occurring within halos in a certain mass range~\cite{Haiman:1999mn,WolcottGreen:2010jp,Nakauchi:2014pma,Visbal:2014fta,Hirano:2015wxa}, and the outcome can also depend on the detailed chemical processes that are included in a simulation~\cite{Glover:2008pz}. Upcoming observations with the James Webb Space Telescope and Square Kilometer Array will be helpful in testing these different scenarios, but in the meantime, there will significant uncertainty in modelling the populations and properties of halos that might undergo \ce{HD} cooling. In the following, we will consider two options--one more optimistic and one more pessimistic---for modelling the halo \ce{HD} luminosity function, and take the difference between the two results as a rough indication of the modelling uncertainty.\footnote{See Ref.~\cite{2020ApJ...888...27N} for an alternative approach to modelling \ce{HD} and \ce{H2} emission from halos at cosmic dawn.}

\subsection{Modelling}
\label{sec:cdmodelling}

To model the mean \ce{HD}(1-0) intensity and power spectrum, we follow the approach of Ref.~\cite{Gong:2012iz}, which computes predictions for the same quantities arising from line emission from \ce{H2} in the halos hosting Pop III stars. This approach also makes use of Eqs.~(\ref{eq:pclus}-\ref{eq_bavg}) and Eq.~\eqref{eq:barI} for computing the power spectrum, but with a different model for halo luminosities than we used for reionization in the previous section.

The halo mass-luminosity relation is written as
\beq
\label{eq:LX}
L_X(M,z) = 4\pi \int_0^{r_{\rm vir}(M,z)} dr\, r^2 n_X(r)
	\sum_{Y}  n_Y(r) \Lambda(X,Y)\ ,
\eeq
where $n_X(r)$ and $n_Y(r)$ are the number densities of $X$ and $Y$ at radius $r$ within the halo and $\Lambda(X,Y)$ is the cooling coefficient for collisions of $Y$ with $X$. The virial radius $r_{\rm vir}$ is given by Eq.~(7) in Ref.~\cite{Gong:2012iz}.  

When evaluating Eqs.~\eqref{eq:barI} and (\ref{eq:pshot}-\ref{eq_bavg}), we follow Ref.~\cite{Gong:2012iz} and set $M_{\rm min}=10\,{\rm M}_\odot$. We also set $M_{\rm max}=10^{13}\,{\rm M}_\odot$ for \ce{HD}(1-0), and use the halo mass function from Ref.~\cite{Tinker:2008ff}. 
We use $r_{\rm min}=1\,{\rm kpc}$ as the lower integration bound in Eq.~\eqref{eq:LX}. The mean intensity and power spectrum integrals are computed with the public \texttt{lim} code\footnote{\url{https://github.com/pcbreysse/lim}}.

The \ce{HD} level populations are dominantly affected by collisions with \ce{H}, \ce{He}, and \ce{H2}. Thus, it remains to model the density profiles of each of these, along with the associated cooling coefficients.

\subsubsection{Density profiles}

The density profiles of each species are related to the total gas profiles, which are observed in several simulations of the first stars (e.g.~\cite{Yoshida:2006bz,McGreer:2008wf,Hirano:2013lba,Hirano:2015wxa}) to follow
\beq
\rho_{\rm gas}(r) = \rho_0 \lp \frac{r}{r_0} \rp^{-2.2}
\eeq
with $r_0$ = $1\,{\rm pc}$ and $\rho_0$ obtained from
\beq
M_{\rm gas} = 4\pi \int_0^{r_{\rm vir}} dr\,r^2\rho_{\rm gas}(r)
\eeq
with $M_{\rm gas} = (\Omega_{\rm b}/\Omega_{\rm m}) M$. ($\rho_{\rm gas}$ thus depends on $M$ and~$z$, but we omit this dependence for brevity.) We can approximate the total gas number density as the sum of the number density of hydrogen and helium nuclei:
\beq
n_{\rm gas}(r) \approx n_{\rm H,nuc}(r)+n_{\rm He,nuc}(r)\ ,
\eeq
where
\begin{align}
n_{\rm H,nuc}(r) &= \frac{f_{\rm H} \rho_{\rm gas}(r)}{m_{\rm H}}\ , \\
n_{\rm He,nuc}(r) &= \frac{(1-f_{\rm H}) \rho_{\rm gas}(r)}{m_{\rm He}}\ ,
\end{align}
$f_{\rm H}=0.739$ is the hydrogen mass fraction, and $m_{\rm H}$, $m_{\rm He}$ are the respective masses. We will ignore helium chemistry and take the number density of \ce{He} atoms to equal that of \ce{He} nuclei, $n_{\ce{He}}(r)=n_{\rm He,nuc}(r)$, while we compute the number density of atomic hydrogen via
\beq
n_{\rm H}(r) = n_{\rm H,nuc}(r) - 2n_{\ce{H2}}(r)\ .
\label{eq:nH}
\eeq
(The $n_{\ce{HD}}/n_{\ce{H2}}$ fraction never exceeds $0.05$ in our model, so we can safely ignore \ce{HD} in Eq.~\ref{eq:nH}.)

For the \ce{H2} and $\ce{HD}$ profiles, we use the results of Ref.~\cite{McGreer:2008wf}, which simulated the evolution of primordial gas clouds starting from cosmological initial conditions, using an adaptive mesh refinement code and including deuterium reactions in their chemical network. In addition to simulations of ``initially unperturbed" (Pop III.1) gas, without any external ionizing sources, they considered an ``initially ionized" (Pop III.2) case in which the gas becomes ionized at $z=20$; although less realistic than a full radiative transfer treatment of the ionization, Ref.~\cite{McGreer:2008wf} argues that this approximation is expected to retain many of the features of a more detailed treatment. In this paper, we use separate models based on the initially unperturbed or ionized simulations, assuming that all halos belong to either one or the other category; these two cases represent pessimistic and optimistic scenarios for the strength of the \ce{HD} intensity mapping signal, with the true signal likely somewhere in between. In the unperturbed case, Ref.~\cite{McGreer:2008wf} found  that halos with $M\lesssim 10^6\msun$ exhibit much stronger \ce{HD} cooling than those with $M\gtrsim 10^6\msun$, and we maintain this distinction in our predictions.

In detail, we take the simulation results for the \ce{H2} mass fraction $X_{\text{H2}}(r)$ from the simulated halos in Ref.~\cite{McGreer:2008wf}\footnote{For the initially unperturbed case, we use the results for Halo1-HD and Halo4-HD from Ref.~\cite{McGreer:2008wf}, taking their \ce{H2} mass fractions $X_{\text{H2}}(r)$, $n_{\ce{HD}}/n_{\text{H2}}$ ratios as a function of $n_{\rm gas}$, and gas temperatures $T(n_{\rm gas})$ to be representative of halos with masses above and below $10^6\msun$ respectively. For the initially ionized case, we take averages of these quantities over the four simulated halos from Ref.~\cite{McGreer:2008wf}, since there are only minor differences between them.}, and translate them into $f_{\text{H2}}$, the number fraction of \ce{H2},  as a function of $r$:
\beq
\label{eq:fH2r}
f_{\text{H2}}(r) \equiv \frac{n_{\text{H2}}(r)}{n_{\rm gas}(r)}
	= \frac{\rho_{\rm gas}(r)}{n_{\rm gas}(r)} \frac{X_{\text{H2}}(r)}{2m_{\rm H}}\ ,
\eeq
using $n_{\rm gas}(r)$ and $\rho_{\rm gas}(r)$ corresponding to the halos for which $X_{\text{H2}}$ was measured. Then, since $n_{\rm gas}(r)$ is a monotonic function, we are able to use Eq.~\eqref{eq:fH2r} to determine $f_{\text{H2}}$ as a function of $n_{\rm gas}$\footnote{This procedure only allows us to determine $f_{\text{H2}}(n_{\rm gas})$ for $n_{\rm gas}<10^{12}\,{\rm cm}^{-3}$ using the results of Ref.~\cite{McGreer:2008wf}. For higher gas densities, we use the directly reported $f_{\text{H2}}(n_{\rm gas})$ values reported in Ref.~\cite{Gong:2012iz}, derived from the simulations in Refs.~\cite{Omukai:2000ic,Yoshida:2006bz}.}. This allows us to compute $n_{\text{H2}}(r)$ for arbitrary halo mass, by inserting the appropriate $n_{\rm gas}(r)$ into
\beq
n_{\text{H2}}(r) = f_{\text{H2}}(n_{\rm gas}[r]) \times n_{\rm gas}(r)\ .
\label{eq:nH2}
\eeq
When necessary, we further split this into profiles of ortho-\ce{H2} and para-\ce{H2}, assuming them to be present in a 3:1 ratio.
We then multiply Eq.~\eqref{eq:nH2} by the measurements of  $n_{\ce{HD}}/n_{\text{H2}}$ (as a function of $n_{\rm gas}$) from Ref.~\cite{McGreer:2008wf} to obtain the \ce{HD} profiles, again using the separate results for initially unperturbed and ionized halos.

\begin{figure}[t]
\begin{centering}
\includegraphics[width=\columnwidth]{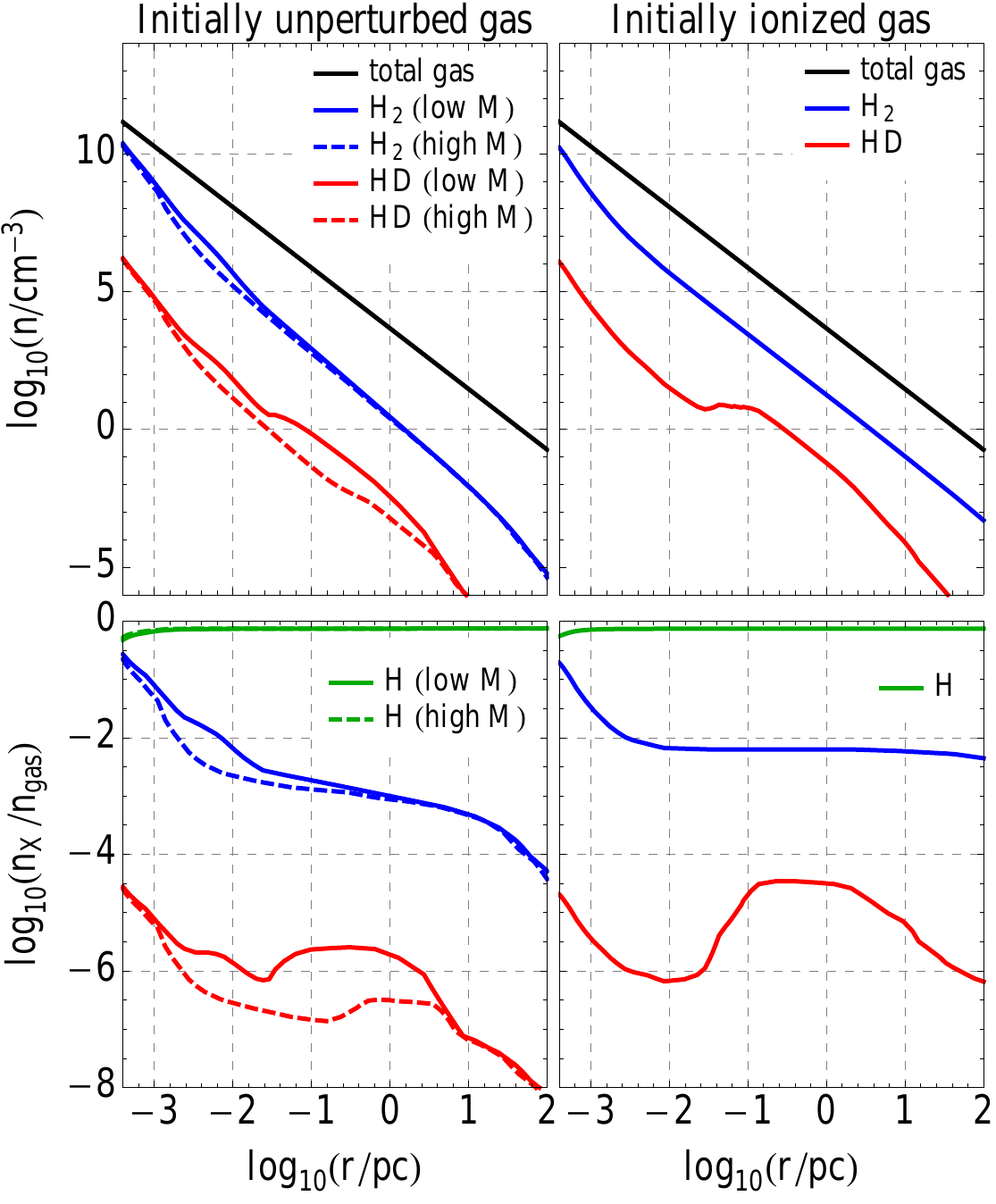}
\caption{\label{fig:profiles}
Number density profiles ({\em upper panels}) and abundances ({\em lower panels}) for $M=10^6\msun$ halos at $z=15$, assuming formation from initially unperturbed gas ({\em left panels}) or initially ionized gas ({\em right panels}), based on the simulations of Ref.~\cite{McGreer:2008wf}. In the initially unperturbed case, we show different models corresponding to high-mass ($M\gtrsim 10^6\msun$) and low-mass ($M\lesssim 10^6\msun$) halos, the latter of which are seen to have higher \ce{H2} and \ce{HD} abundances in simulations. The initially unperturbed and ionized cases represent optimistic and pessimistic scenarios for modelling the HD intensity mapping signal from cosmic dawn.
}    
\end{centering}
\end{figure}

The upper panels of Fig.~\ref{fig:profiles} show the predicted number density profiles of total gas, \ce{H2}, and \ce{HD} for each type of halo in our model (initially unperturbed gas with $M<10^6\msun$ and $>10^6\msun$, and initially ionized gas), while the lower panels show the chemical abundances with respect to the total gas number density. These curves reflect the increased production of \ce{H2} and \ce{HD} for lower-mass initially-unperturbed halos, due to the generally lower temperatures in these halos as compared to those of higher mass, and the much larger production of these species when the gas is initially ionized, catalyzed by the larger abundance of free electrons~\cite{McGreer:2008wf}.

\subsubsection{Cooling coefficients}

Following Refs.~\cite{Gong:2012iz,Hollenbach:1979}, for an optically thin line, the cooling coefficient $\Lambda(X,Y)$ can be written as
\beq
\Lambda(X,Y) = \frac{\Lambda_{\rm LTE}(X,Y)}
	{1+n_{\rm cr}(Y)/n_Y}\ ,
\label{eq:LambdaXY}
\eeq
where $\Lambda_{\rm LTE}(X,Y)$ is the cooling coefficient at local thermal equilibrium and $n_{\rm cr}(Y)$ is the critical density of $Y$ required to reach LTE. The cooling coefficient at LTE is
\beq
\Lambda_{\rm LTE}(X,Y) = \frac{1}{n_Y} A_{JJ'} \frac{g_J}{g_{J'}} 
	e^{-\frac{\Delta E_{JJ'}}{k_{\rm B}T}} \Delta E_{JJ'}\ ,
	\label{eq:LambdaLTE}
\eeq
where $A_{JJ'}$ is the Einstein coefficient for the transition of interest, and our notation assumes that this transition is between two rotational levels $J$ and $J'$. For \ce{HD}(1-0), $A_{10} \approx 5.1\times 10^{-8}\,{\rm s}^{-1}$~\cite{Flower2000}, $g_1=3$, $g_0=1$, and $\Delta E_{10}/k_{\rm B}=128\,{\rm K}$. 
We can approximate
\beq
\frac{n_{\rm cr}(Y)}{n_Y} \approx \frac{ \Lambda_{\rm LTE}(X,Y) }{ \Lambda_{n\to 0}(X,Y) }\ ,
\eeq
where the low-density cooling coefficient $\Lambda_{n\to 0}(X,Y)$ is
\beq
\Lambda_{n\to 0}(X,Y) = \kappa_{JJ'}(X,Y;T) \frac{g_J}{g_{J'}} 
	e^{-\frac{\Delta E_{J\to J'}}{k_{\rm B}T}} \Delta E_{J\to J'}
	\label{eq:Lambdan0}
\eeq
with $\kappa_{JJ'}(X,Y;T)$ the collisional de-excitation coefficient for $J\to J'$ in $X$ in collisions with $Y$ at temperature~$T$. We obtain the gas temperature profiles $T(r)$ needed for Eqs.~\eqref{eq:LambdaLTE} and~\eqref{eq:Lambdan0} by taking the $T(n_{\rm gas})$ measurements from the simulations in Ref.~\cite{McGreer:2008wf} and evaluating them on the $n_{\rm gas}(r)$ profiles for each halo mass we consider.

For \ce{HD} colliding with \ce{H}, \ce{He}, and \ce{H2}, we use the collisional coefficients from Ref.~\cite{Flower2000}\footnote{Available at \url{http://ccp7.dur.ac.uk/cooling_by_HD/}.}, which are based on fits to the original computations in Refs.~\cite{FlowerRoueff:1999,RoueffZeippen:1999}. For the 1-0 transition we are concerned with here, these rates are in excellent agreement with more recent computations~\cite{Nolte:2012,Balakrishnan:2018,Desrousseaux:2018}.

\begin{figure}[t]
\begin{centering}
\includegraphics[width=\columnwidth]{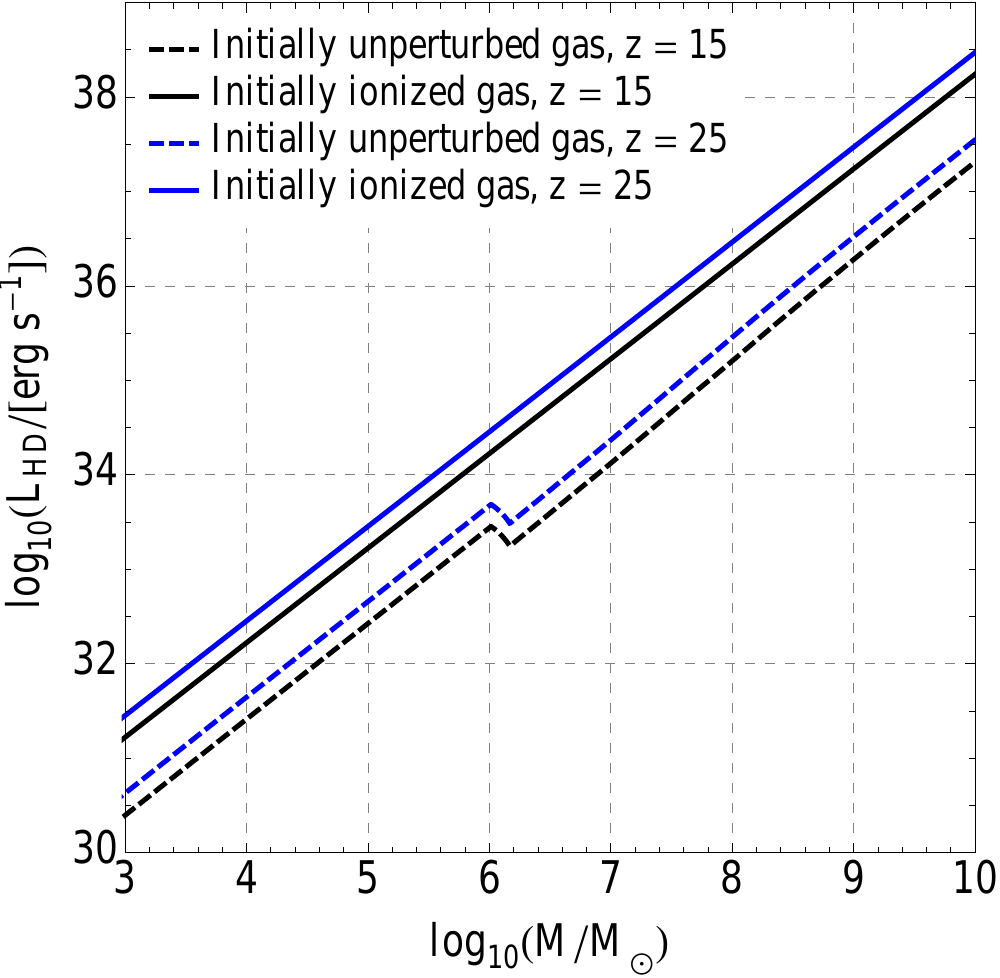}
\caption{\label{fig:Lmz}
Relationship between \ce{HD}(1-0) luminosity and halo mass in our model for cosmic dawn, for initially unperturbed ({\em dashed}) and initially ionized ({\em solid}) gas clouds, and at two representative redshifts. This relationship is normalized higher in the initially unperturbed case for halos with $M\lesssim 10^6\msun$ than for higher masses, and is much higher for initially ionized halos, reflecting the increased \ce{HD} abundance in those cases.
}    
\end{centering}
\end{figure}

Using the ingredients described above, the final $L_{\ce{HD}}(M,z)$ relations we compute in the initially unperturbed and ionized cases are shown in Fig.~\ref{fig:Lmz} at two different redshifts. As expected, we see much higher luminosities in the initially ionized case, due to the higher abundance of \ce{HD}, as well as higher luminosities for initially unperturbed halos with $M\lesssim 10^6\msun$. The luminosities are higher at higher redshift, because a halo with a given mass is denser at earlier times (we use virial halo masses in our model, and the virial radius for a given mass is lower at earlier times). We show the analogous plot for $L_{\rm{H2}}(M,z)$ in Appendix~\ref{app:h2}.

\subsubsection{Consistency Check}

\begin{figure}
\centering
\includegraphics[width=\columnwidth]{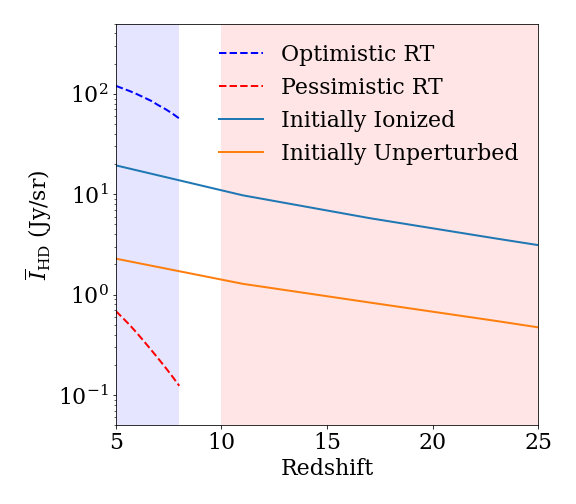}
\caption{Consistency check between the reionization-era models presented in Section \ref{sec:reion} (dashed curves) and the cosmic dawn-era models presented in Section \ref{sec:cd} (solid curves).  The redshift ranges where we apply the models are shaded, with the cosmic dawn models extrapolated to lower redshifts for comparison.  We note that the initially ionized (cyan) and initially unperturbed (orange) models lie in the range of the optimistic (blue) and pessimistic (red) radiative transfer models. 
}
\label{fig:consistency}
\end{figure}

The radiative transfer model derived in Sec.~\ref{sec:reion} can only be computed out to redshift $z\lesssim8$, as that is the limit of the simulations presented in Ref.~\cite{Lagos2011}.  We consider the models given in this section to be most accurate at $z\gtrsim 10$, because the ionizing background at lower redshifts will likely be strong enough to affect the \ce{HD} abundances in the relevant halos in a way not captured by our models.  However, we can still extrapolate our initially ionized and unperturbed models to check consistency with the previous section, though we will not use this extrapolation to make forecasts.  Fig.~\ref{fig:consistency} shows the results of this extrapolation.  We can clearly see that, when extended to the reionization era, both the ionized and unperturbed models give mean intensities within the range of radiative transfer models considered above.  Thus, though our models are highly uncertain, they do show an encouraging level of consistency.

\subsection{Detectability of cosmic dawn signal}

\begin{figure*}
\centering
\includegraphics[width=\textwidth]{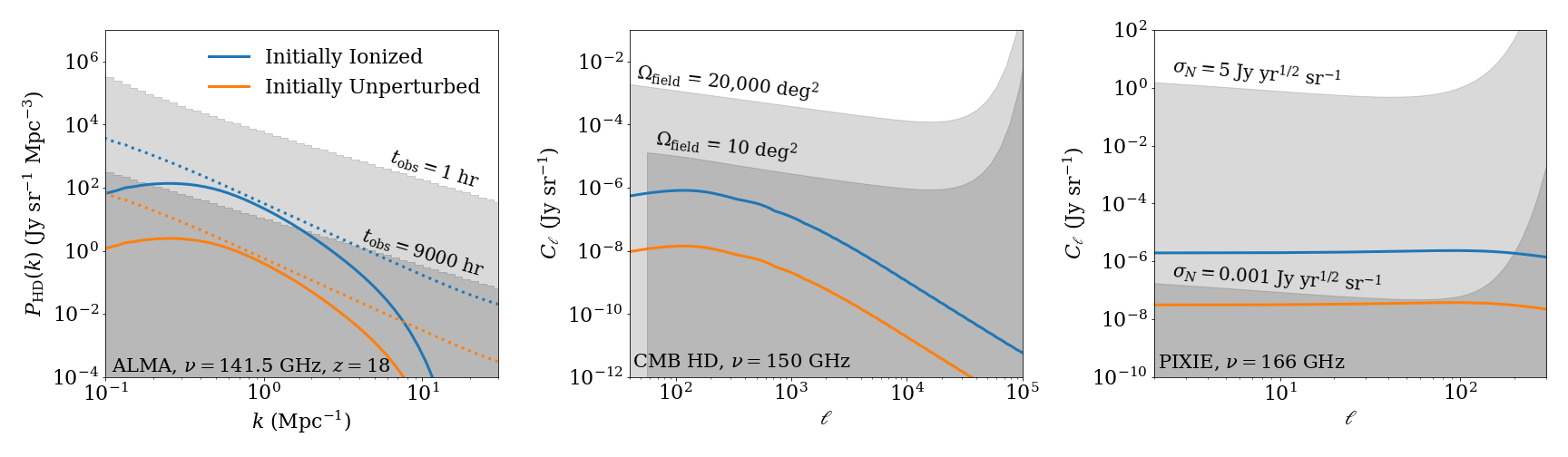}
\caption{
Forecasted signals and sensitivities for cosmic-dawn era HD intensity mapping measurements. Power spectra (solid lines) are shown for the initially ionized (blue) and unperturbed (orange) models.  {\it Left panel:} ALMA Band 4 observations over a 0.01 deg$^2$ field.  Dotted lines show power spectra without instrumental resolution or finite-survey effects.  The light gray band is the power spectrum error in bins of $\Delta\log k=0.04$ for a single 1 hr pointing, the dark grey shows the same for a 9000 hr observation.  The latter is necessary to obtain a 5$\sigma$ detection of the ionized model.  {\it Center panel:} CMB HD observations, plotted as angular power spectra due to the large solid angle coverage.  The light grey band shows the error for the planned $\sim$half sky survey area, the dark band assumes the full observing time is spent on a 10 deg$^2$ field.  The latter only obtains an overall SNR of $\sim1$.  {\it Right panel:} Space-based observations with a PIXIE-like instrument.  The light grey band uses the planned PIXIE sensitivity, while the dark grey uses the substantially increased sensitivity from Ref.~\cite{Abitbol:2019ewx} to obtain a strong detection of both models. 
}
\label{fig:cosmic_dawn}
\end{figure*}

The HD(1-0) line from cosmic dawn falls conveniently into a frequency range observed by a variety of existing and planned cosmological measurements.  We will demonstrate here that, though the signal modeled above is too faint for currently-planned surveys, it may be detectable with future instruments.

In this section, we forecast the (purely statistical) detection significance of the HD(1-0) power spectrum in the two models discussed above (initially unperturbed vs.\ initially ionized).  These two models serve as an extremely rough illustration of the uncertainty in the HD modeling at cosmic dawn, though of course the possibility remains for a significantly brighter or fainter signal.  Under our current models, we find that current experiments cannot reasonably obtain signal-to-noise ratios better than $\mathcal{O}(10^{-2})$, putting detection out of reach for the near term.  This is partially due to the essentially low amplitude of the signal, but also because existing observations are poorly-optimized for the HD observation. 

Below we discuss the pros and cons of several existing and planned instruments for HD measurements, and discuss what it would take in each case for a detection.  For details on the modelling of each experiment, see Appendix \ref{app:cdforecasts}. We remind the reader that we only consider statistical significance here, assuming that all systematics can be perfectly mitigated, so even seemingly hopeful SNR values will require further study to take various systematics into account.

\subsubsection{ALMA}

The cosmic-dawn era HD(1-0) line falls into ALMA bands 4 and 5, with the two bands spanning from redshift $z\sim11-20$.  ALMA has excellent overall sensitivity, and because it is an interferometer it has extremely high angular resolution, allowing for very deep surveys of small patches.  ALMA has in fact already proven a capable tool for small-scale intensity mapping surveys \cite{Keating2020}, detecting the aggregate emission from several unresolved CO rotational transitions in band 3.  The left panel of Fig.~\ref{fig:cosmic_dawn} shows ALMA sensitivities for a hypothetical HD survey centered at 142 GHz, or $z\sim14$.  Sensitivities assume a single 7.5 GHz frequency slice over a total survey area of 0.01 deg$^2$, chosen to optimize the trade-off between instrument noise and sample variance error.

However, ALMA has two key drawbacks which make it less-suited to detecting cosmic-dawn HD.  First, its high resolution is limited to very small survey areas.  This means that it is primarily sensitive to smaller scales in the power spectrum (in other words, the large-scale cutoff in $W(k)$ from survey area appears at relatively high~$k$).  The HD power spectrum has very little shot noise at these redshifts (c.f. the dotted unconvolved power spectra in Fig.~\ref{fig:cosmic_dawn}), so the signal is quite low on these scales.  Second, and more crucially, ALMA is not a dedicated survey instrument like the others we will consider in this section.  For a 5$\sigma$ detection of our brighter model, we would need to survey our field for $\sim9000$ hours.  This is not an unreasonable survey time for a dedicated CMB instrument, but it is completely unrealistic for an instrument as subscribed as ALMA.  

\subsubsection{Ground-based CMB instruments}

Cosmic-dawn HD emission also falls into the frequency range of many CMB surveys.  Unlike ALMA, these instruments are generally single-purpose survey projects, potentially allowing them to reach the depths required to detect the faint signal we model here.  However, as a trade-off, these experiments do not have the same raw sensitivity of the ALMA interferometer.  

The proposed CMB-HD experiment~\cite{Sehgal:2019ewc} is likely to be the best of the near-future ground-based surveys for our purposes, due to its low noise levels and high angular resolution (and because of its auspicious name).  The proposed survey would cover roughly half of the sky for 7.5 years.  Because the survey area is so large, the implicit flat-sky approximation we have used to compute 3D power spectra thus far is inadequate.  In addition, CMB-HD (and most CMB surveys) do not have the spectral resolution of dedicated intensity mapping surveys, so the third dimension is largely unnecessary.   We will instead forecast the 2D angular power spectrum $C_\ell$ for CMB-HD, which can be related to the 2D spectrum discussed thus far by
\beq
C_\ell = \frac{2}{\pi}\int dk k^2P_{\rm{HD}}(k)\left[\int drf(r)j_\ell(kr)\right]^2,
\eeq
where $j_\ell$ is the spherical Bessel function and $f(r)$ is the radial window function of the survey.

For the CMB-HD frequency channel centered at 150~GHz, we obtain a signal-to-noise ratio of $0.02$ integrated over all scales for the brighter of our two models.  This is sadly still too low to be useful, and once again the primary issues stem from the fact that CMB-HD is not optimized for the HD measurement.  Because the HD signal is so faint, a half-sky survey is much larger than ideal.  If we assume the same survey time is spent on a 10 deg$^2$ field, the SNR rises to 0.75, which is still low, but a substantial improvement.  

The other challenge comes from the lack of spectral resolution.  The CMB is a 2D surface with smooth frequency evolution, so most experiments are designed around a few very wide frequency bands.  The intensity mapping signal, however, is fully three-dimensional, so many modes are lost with only wide frequency bands.  On top of this, the primary means of removing foreground emission from an intensity map is to make use of the frequency structre of the signal.  Though we do not directly model foregrounds here, this would provide another significant challenge for a broad-band survey.

If we make the rough approximation that SNR scales with the square root of the number of frequency channels, then a version of CMB-HD with the bands split into $\mathcal{O}(50)$ channels could reach an SNR of 5 for the brighter model.  Since increasing the frequency resolution is a much more substantial hardware change than simply varying survey area, we do not attempt a full forecast for this hypothetical modification of CMB-HD.

\subsubsection{CMB satellites}

Orbiting CMB experiments have much more freedom in terms of frequency coverage than their ground-based equivalents due to the absence of atmospheric contamination.  It is thus natural to examine whether any of these could detect HD as well.

Unfortunately, despite their advantages, most CMB satellites fall victim to the same difficulties as CMB-HD.  Surveys optimized for CMB observations tend to cover large sky areas, creating very high noise levels for our faint HD signal.  Also, again due to the broad-band nature of CMB emission, most surveys do not have particularly high frequency resolution.  We performed forecasts for several current and proposed CMB satellites, including Planck \cite{Akrami:2018vks}, LiteBIRD \cite{HazumiLitebird}, and PICO \cite{Hanany:2019lle}.  Of these surveys, the current-best-estimate PICO setup gave the best SNR at cosmic dawn, rising to $0.002$ for their 155 GHz channel for the initially ionized model.  As with CMB-HD, this could potentially be improved by narrowing the survey area and adding frequency resolution, but not without dramatically altering the basic nature of the survey.

There is another proposed survey, however, which has a quite different design.  The PIXIE survey \cite{Kogut2011} has a goal of measuring distortions to the blackbody shape of the CMB spectrum.  Because of this, it has much finer frequency resolution than a typical survey, making it much better suited for intensity mapping measurements \cite{Switzer2017}.  This extra resolution however comes at the cost of lower overall sensitivity.  An all-sky PIXIE map spanning 15~GHz centered at 166 GHz obtains an extremely low SNR of $\sim10^{-7}$, based on the sensitivities quoted in Ref.~\cite{Kogut2011}.  Future, more advanced versions discussed in the literature could do better.  For example, the enhanced version of the PIXIE experiment described in Ref.~\cite{Abitbol:2019ewx}, which is designed to observe the time-evolution of the CMB blackbody, has enough sensitivity to detect even the fainter of our two models at high significance. This statement assumes an identical survey to the baseline from Ref.~\cite{Kogut2011}, but with a 5000-fold increase in depth. Furthermore, this futuristic experiment would also likely have the statistical power to see the \ce{HD}(1-0) {\em monopole}, since the range of our predictions for $\bar{I}_{\ce{HD}}$ from cosmic dawn ($\mathcal{O}(1)$ to $\mathcal{O}(10)$ Jy$\,$sr$^{-1}$; see Fig.~\ref{fig:consistency}) are well above the monopole sensitivities computed for this experiment in Ref.~\cite{Abitbol:2019ewx} ($\mathcal{O}(10^{-3})$ to $\mathcal{O}(10^{-2})$ Jy$\,$sr$^{-1}$; see their Figure~2). However, separation from other foregrounds would pose a significant challenge for such a measurement.

\subsubsection{Cross-correlations}

Throughout the history of cosmology, one of the most effective ways to detect a faint signal has been to cross-correlate with an additional large-scale structure tracer.  Cross-correlations can improve signal-to-noise, and serve as an excellent tool for isolating a signal at a specific redshift from contaminating foregrounds. At the extreme redshifts considered here, there are not likely to be direct-imaging catalogs available for the foreseeable future, so the only likely target for cross-correlations will be with other intensity mapping surveys.  As cosmic dawn by definition occurs before significant star formation has taken place, there are very few emitting species to cross-correlate with.  Two possible cross-correlation targets exist: the 21$\,$cm hydrogen spin-flip transitions, and transitions from ordinary molecular hydrogen.

Across all redshifts, the 21$\,$cm line is the most common target for intensity mapping.  The HERA interferometer, currently operating in South Africa, has the ability to observe 21$\,$cm photons from as far as redshift 20 \cite{DeBoer2017}, thus allowing for overlap with our hypothetical HD observations.  In addition to helping detect HD, such a cross-correlation could provide significant science benefits to both surveys in the form of reduced foregrounds.  Foregrounds are an immense challenge for 21$\,$cm observations, often several orders of magnitude brighter than their signal \cite{Liu2009}.  For HD, a 21$\,$cm correlation could make up for the lack of redshift resolution in a broadband CMB survey, similar to what was done in Refs. \cite{Pullen2018,Yang2019}.

A 21$\,$cm-HD correlation would still likely require at minimum a more sensitive HD measurement than is currently possible, along the lines of the discussion in the previous section.  HERA is targeting a first detection of high-redshift 21 cm, and is unlikely to obtain the kind of extremely high SNR that would be necessary to bring out the faint HD line.

Ref. \cite{Gong:2012iz} proposed using intensity maps of H$_2$ transitions to map cosmic dawn.  Though H$_2$ lacks the slight asymmetry that gives rise to rotational transitions in HD, its much greater abundance may lead to a signal bright enough to detect nonetheless.  Since the two lines will both come from early molecular clouds, an H$_2$ correlation would provide an excellent cross-check on an HD observation.  The brightest H$_2$ line from cosmic dawn falls into the frequency range of the proposed Origins Survey Spectrometer \cite{Cooray:2019ivm}.  Unfortunately, though Ref. \cite{Gong:2012iz} argued that a near-future experiment would have the potential to detect H$_2$, we find that an updated forecast with currently planned surveys does not have sufficient sensitivity (see Appendix \ref{app:h2}).  Thus, while an HD-H$_2$ correlation is promising in principle, we would likely need mildly futuristic observations of both lines to reach cosmic dawn.

\section{Dark Ages}
\label{sec:darkages}

Finally, in this section we ask whether \ce{HD} could be used for intensity mapping from the so-called dark ages ($z\gtrsim 30$), before the first stars formed. The most promising way to access this era of cosmic history is 21$\,$cm intensity mapping~\cite{Loeb:2003ya,Burns:2019zia,Furlanetto:2019jso}, which will be rather difficult on its own, but given the amount of pristine cosmological information available during this epoch, it is worthwhile to explore other possible probes. Previously explored options for intensity mapping include hyperfine transitions in deuterium ($\lambda\approx92\,{\rm cm}$), which could provide a measurement of the primordial [D/H] ratio if 92$\,$cm maps are cross-correlated with 21$\,$cm maps, although this will be difficult in practice~\cite{Sigurdson:2005mp}; or in ${}^3$\ce{He+} ($\lambda\approx 3.5\,{\rm cm}$), which will be essentially invisible during the dark ages due to the lack of contrast between the spin temperature in the IGM and the CMB temperature~\cite{Bagla:2009mx,McQuinn:2009ng}. 

At $z\lesssim 60$, \ce{HD} will be present in the IGM with $[\ce{HD}/\ce{H}]\approx 4\times10^{-10}$~\cite{Galli:2012rf}; while this is tiny compared to the abundance of \ce{H}, the spontaneous decay rate for \ce{HD}(1-0) ($A_{10} \approx 5.1\times 10^{-8}\,{\rm s}^{-1}$) is seven orders of magnitude higher than for the 21$\,$cm transition ($A_{10}\approx2.9\times10^{-15}\,{\rm s}^{-1}$). Furthermore, the signal would fall in the range between $40$ and $90\,{\rm GHz}$, which is already targeted by CMB experiments and will be much less impacted by the galactic synchrotron and ionospheric effects that will pose major obstacles for 21$\,$cm measurements from the dark ages (although there will be other bright continuum foregrounds, most notably the CMB blackbody itself). With this in mind, we explore the \ce{HD} case in more detail.

We wish to estimate the mean brightness temperature arising from \ce{HD}(1-0) from the IGM, in contrast with the backlight CMB. This is given by (e.g.~\cite{Furlanetto:2006jb})
\beq
T_{\rm b}(z) = \frac{T_{\rm ex}(z)-T_\gamma(z)}{1+z} (1-e^{-\tau})\ ,
\label{eq:tb}
\eeq
where $T_\gamma$ and $T_{\rm ex}$  are the CMB temperature and the excitation temperature for the HD transition, respectively, and $\tau$ is the associated optical depth. This formula holds even outside of the Rayleigh-Jeans regime, as long as $T_{\rm ex}$ is very close to $T_\gamma$ (which we will find to be the case here). The excitation temperature is defined by
\beq
\frac{n_1(z)}{n_0(z)} = \frac{g_1}{g_0} e^{-T_{10}/T_{\rm ex}(z)}\ ,
\label{eq:texdef}
\eeq
where $n_1$ and $n_0$ are the number densities of the upper and lower states, and $T_{10} =128\,{\rm K}$. The optical depth can be written as~\cite{Sigurdson:2005mp}
\beq
\tau = \frac{g_1}{g_1+g_0} \frac{1}{8\pi} \frac{hc}{k_{\rm B} T_{\rm ex}(z)}
	\frac{A_{10}}{H(z)} \lambda_{10}^2 n_{\ce{HD}}(z)\ ,
	\label{eq:tau}
\eeq
where $H(z)$ is the Hubble parameter, assuming that the dominant source of line broadening is the Hubble flow.

Standard expressions for the excitation temperature of \ce{H} cannot be directly applied to \ce{HD}, because the Rayleigh-Jeans approximation for the blackbody intensity is not valid in this case. We must instead work forward from the equation for detailed balance of the  upward and downward transitions~\cite{Furlanetto:2006jb},
\beq
n_1 \lb C_{10} + A_{10} + B_{10} I_\gamma \rb
	= n_0 \lb C_{01} + B_{01} I_\gamma \rb\ ,
	\label{eq:eqb}
\eeq
where $C_{10}$ is the collisional de-excitation rate of \ce{HD}.\footnote{For 21$\,$cm at lower redshift, one must also account for the coupling between \ce{H} and the UV radiation background (the so-called Wouthuysen-Field effect~\cite{Wouthuysen:1952,Field:1958}). We are working at high enough redshift that we can ignore any possible analogous effect for \ce{HD}.}
The Einstein $B$ coefficients are related to $A_{10}$ by
\beq
B_{10} = \frac{c^2}{2h\nu^3} A_{10} 
\ , \quad
	B_{01} = \frac{g_1}{g_0} B_{10}\ ,
	\label{eq:einsteinB}
\eeq
while $C_{01}$ and $C_{10}$ are related by
\beq
\frac{C_{01}(z)}{C_{10}(z)} = \frac{g_1}{g_0} e^{-T_{10}/T_{\rm K}(z)}\ ,
\eeq
where $T_{\rm K}$ is the gas temperature, which we approximate as $T_{\rm K}\approx 0.02(1+z)^2\,{\rm K}$~\cite{Galli:2012rf}. The rate $C_{10}$ is a sum over particle species,
\beq
C_{10} = \sum_Y n_Y \kappa_{10}(\ce{HD},Y;T)\ ,
\eeq
where $n_Y$ is the number density of species $Y$ and $\kappa_{10}$ is the same rate coefficient that appeared in Eq.~\eqref{eq:Lambdan0}. We consider collisions of \ce{HD} with \ce{H} and \ce{He} in our numerical calculations, \ce{H2} having a negligible effect. The radiation (CMB) intensity is a blackbody with temperature $T_\gamma$,
\beq
I_\gamma(\nu,z) = \frac{2h\nu^3}{c^2} \frac{1}{\exp\lb h\nu/k_{\rm B} T_\gamma(z)  \rb - 1}\ ,
\label{eq:igamma}
\eeq
and we evaluate it and Eq.~\eqref{eq:einsteinB} at a frequency corresponding to $T_{10}$. We use $T_\gamma(z)=2.725(1+z)\,{\rm K}$ for the CMB temperature.

By combining Eqs.~\eqref{eq:texdef} and~\eqref{eq:eqb}-\eqref{eq:igamma}, we can solve for $T_{\rm ex}(z)$, and use the result in Eqs.~\eqref{eq:tau} and~\eqref{eq:tb} to obtain the mean brightness temperature. The result is that $T_{\rm b}(z)$ for \ce{HD}(1-0) reaches a maximum of $\sim 10^{-15}\,{\rm K}$ at $z=60$. As points of reference, the 21$\,$cm brightness temperature reaches a maximum (in absorption) of roughly $50\,{\rm mK}$ during the dark ages, with the corresponding number for the 92$\,$cm \ce{D} line being of order $\mu$K~\cite{Sigurdson:2005mp,Furlanetto:2006jb}.

There are a few physical effects that act to suppress~$T_{\rm b}$ down to such a small level. First, the excitation temperature remains tightly coupled to the CMB temperature in this epoch. For 21$\,$cm, collisions drive the \ce{H} excitation temperature towards the gas temperature at $30\lesssim z \lesssim 150$. For \ce{HD}, however, stimulated transitions by the CMB are much more efficient (due to the higher Einstein $A$ value), while collisional excitation by the gas is much less efficient, due to the greater energy required ($\Delta E/k_{\rm B}=128\,{\rm K}$, versus $\Delta E/k_{\rm B}=0.068\,{\rm K}$ for 21$\,$cm). Numerically, this implies that $T_{\rm ex}$ does not differ from $T_{\rm \gamma}$ by more than $10^{-3}\,{\rm K}$ at these redshifts. Second, despite the higher Einstein $A$ value, the optical depth for these transitions is suppressed by the low \ce{HD} abundance and shorter wavelength compared to 21$\,$cm, resulting in a value of $\tau\approx 5\times 10^{-11}$ at $z=60$ (versus $\tau\approx 0.04$ for 21$\,$cm).

Furthermore, when translating the $\mathcal{O}(10^{-15}\,{\rm K})$ brightness temperature to an intensity at a desired observing frequency (e.g. $\nu=45\,{\rm GHz}$, or $T\approx 2.1\,{\rm K}$, for $z=60$), one finds that the result is deep in the Wien tail of the blackbody distribution, implying an observed intensity that is suppressed by $\exp(-h\nu/k_{\rm B}T_{\rm b})$. This firmly places the signal beyond the reach of any conceivable experiment.

\section{Conclusion}
\label{sec:conclusion}

Reionization and cosmic dawn represent key unobserved frontiers in modern cosmology.  We have discussed here a novel potential window into the high-redshift galaxies that drive reionization, the molecular clouds which birthed the earliest stars, and the intergalactic medium during the cosmic dark ages.  While hydrogen intensity mapping is likely to remain the primary tracer of large-scale structure at extreme redshifts, HD could in principle represent a powerful complement to commonly discussed 21$\,$cm intensity maps.  

We have presented HD models covering reionization ($z\sim6-10$), cosmic dawn ($z\sim10-30$), and the dark ages ($z\gtrsim30$).  Of these regimes, we find (not unexpectedly) that the epoch of reionization represents the easiest target.  While current [CII] intensity mapping experiments lack the sensitivity needed to detect HD, modest future improvements should bring the signal with reach through internal cross-correlations.  Future designs should possess the statistical power to detect even our more pessimistic models (subject to appropriate control of systematics), adding HD to our ever-growing toolbox of reionization observations.

At cosmic dawn, where there are far fewer other observables available, we find that the HD signal is, while present, too weak to be detected by current experiments.  There are many observatories designed for other purposes which cover the relevant frequencies, but none of them can detect even our most optimistic signal.  Our forecast is less pessimistic for the near-to-moderate future however, with modest evolutions of current surveys able to begin cutting into the range of possible models.  

Some hypothetical HD-sensitive instruments could include:
\begin{itemize}
\item An ALMA-like observatory capable of dedicating $\sim$years of observing time to a deep intensity mapping survey,
\item A survey with CMB-HD-level sensitivity spent on a $\sim$few deg$^2$ field, with modest spectral resolution, or
\item A future, more sensitive spectroscopic CMB satellite, for example the enhanced version of the PIXIE satellite presented in Ref.~\cite{Abitbol:2019ewx}.
\end{itemize}

The first two of these in particular are easily within the realm of present technology, though it may be some time before a dedicated HD measurement would justify the expense.  The latter example requires more advancement, but would allow an HD measurement as part of a greater range of science goals.  One could further improve the strength of an HD measurement by cross-correlating with either 21 cm or H$_2$ intensity maps, though future instruments would likely be necessary for these lines as well.

Finally, we studied the possibility of detecting a global IGM HD signal during the dark ages.  As HD lacks the strong collisonal coupling that gives rise to a 21 cm dark ages signal, we find that there is no reasonably detectable signal from the pre-cosmic-dawn era.  The first molecular clouds at $z\sim15-30$ thus represent the earliest detectable source of HD emission.

\begin{acknowledgements}

We thank Matt Bradford, Yan Gong, Mat Madhavacheril, Margot Mandy, David Neufeld, Alexander van Engelen, and Dongwoo Chung for useful discussions. Research at Perimeter Institute is supported in part by the Government of Canada through the Department of Innovation, Science and Industry Canada and by the Province of Ontario through the Ministry of Colleges and Universities. JM is supported by the US Department of Energy under grant no.~DE-SC0010129. This research was undertaken, in part, thanks to funding from the Canada Research Chairs Program. LCK acknowledges the support of a Beatrice and Vincent Tremaine Fellowship.

\end{acknowledgements}

\appendix
\section{Mass-Luminosity Scaling}
\label{app:MofM}

When estimating power spectra for reionization-era \ce{HD} from our radiative transfer models, we made the simplifying assumption that the HD luminosity of a halo scales linearly with its mass.  We will show here that the potential error introduced due to this assumption is small compared to the other modelling uncertainties.  

Our model gives an estimate for the sky-averaged mean intensity of the HD(1-0) line based on assumptions about the conditions within reionization-era molecular clouds (Eqs.~\ref{eq:lhd}-\ref{eq:barI}).  We then need to decide how the HD emission is distributed among halos.  To check the relative importance of this choice, we will compute power spectra for our ``optimistic" and ``pessimistic" models assuming five different halo mass scalings:

\begin{itemize}
\item The simple linear halo mass scaling used above.
\item The same linear scaling, but with $M_{\rm{min}}$ dropped to $10^6\ M_{\odot}$ from $10^8\ M_{\odot}$.
\item The scaling used for CO emission from Ref. \cite{Li:2015gqa}, based on the star formation histories of Ref.~\cite{Behroozi2013}.
\item The same CO scaling, but with the 0.3 dex lognormal scatter between luminosity and halo mass used in Ref.~\cite{Li:2015gqa} included.
\item The scaling used for [CII] emission at reionization from Ref.~\cite{Silva2015}.  We use their model ``m2" as it is said to be the median of the four models presented.
\end{itemize}

\begin{figure}
\centering
\includegraphics[width=\columnwidth]{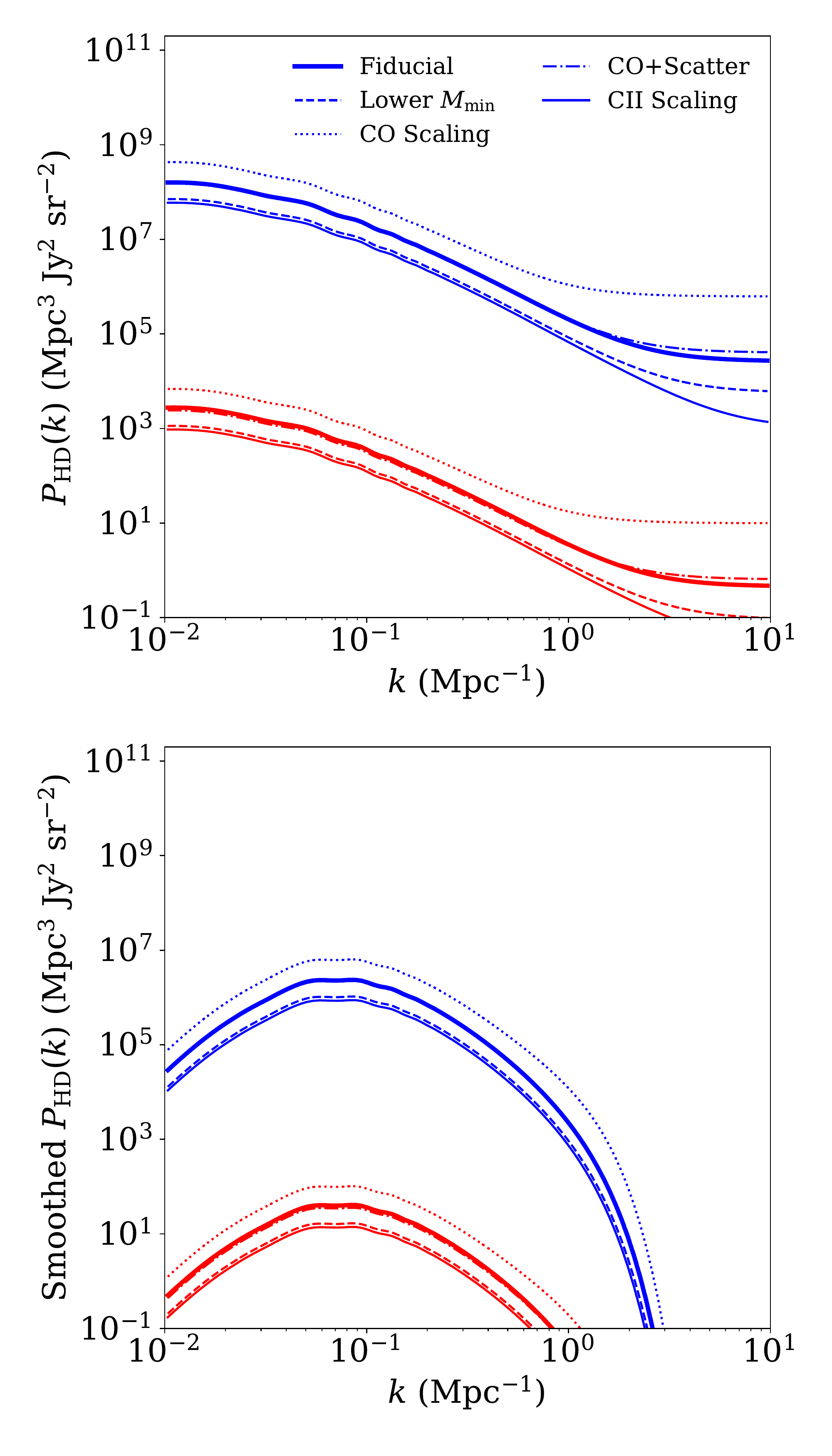}
\caption{Effect of the shape of the HD $L(M)$ relation on the power spectrum, both without (bottom) and with (top) instrumental effects included.  The fiducial model linear model is plotted as the thick solid line, other possibilities include the same model with $M_{\rm{min}}$ reduced to $10^6\ M_\odot$ (dashed), the CO model from Ref. \cite{Li:2015gqa} with (dot-dashed) and without (dotted) scatter included, and the [CII] model from Ref. \cite{Silva2015} (thin solid). Curves are plotted for the optimistic (blue) and pessimistic (red) radiative transfer models.}
\label{fig:MofM}
\end{figure}

While these scalings certainly do not capture the full range of possible behaviors, they do cover a wide range of possibilities.  For each model, we enforce the overall mean normalization computing using Eqs.~(\ref{eq:lhd}-\ref{eq:barI}), then predict the updated bias and shot noise.  The results are shown in Fig.~\ref{fig:MofM} both with and without resolution effects included.  We find that there can indeed be substantial differences between these models, particularly in the shot-noise regime.  However, this difference does not outweigh the difference in mean intensities between the two models.  On the scales we are sensitive to for FYST-like experiments, the effects of the mass scaling are entirely subdominant.

\section{\ce{H2} intensity mapping at cosmic dawn}
\label{app:h2}

In this appendix, we consider intensity mapping of \ce{H2} at cosmic dawn using the 5-3 rotational transition, identified in Ref.~\cite{Gong:2012iz} to be the most promising \ce{H2} transition for detection.

\begin{figure}[t]
\begin{centering}
\includegraphics[width=\columnwidth]{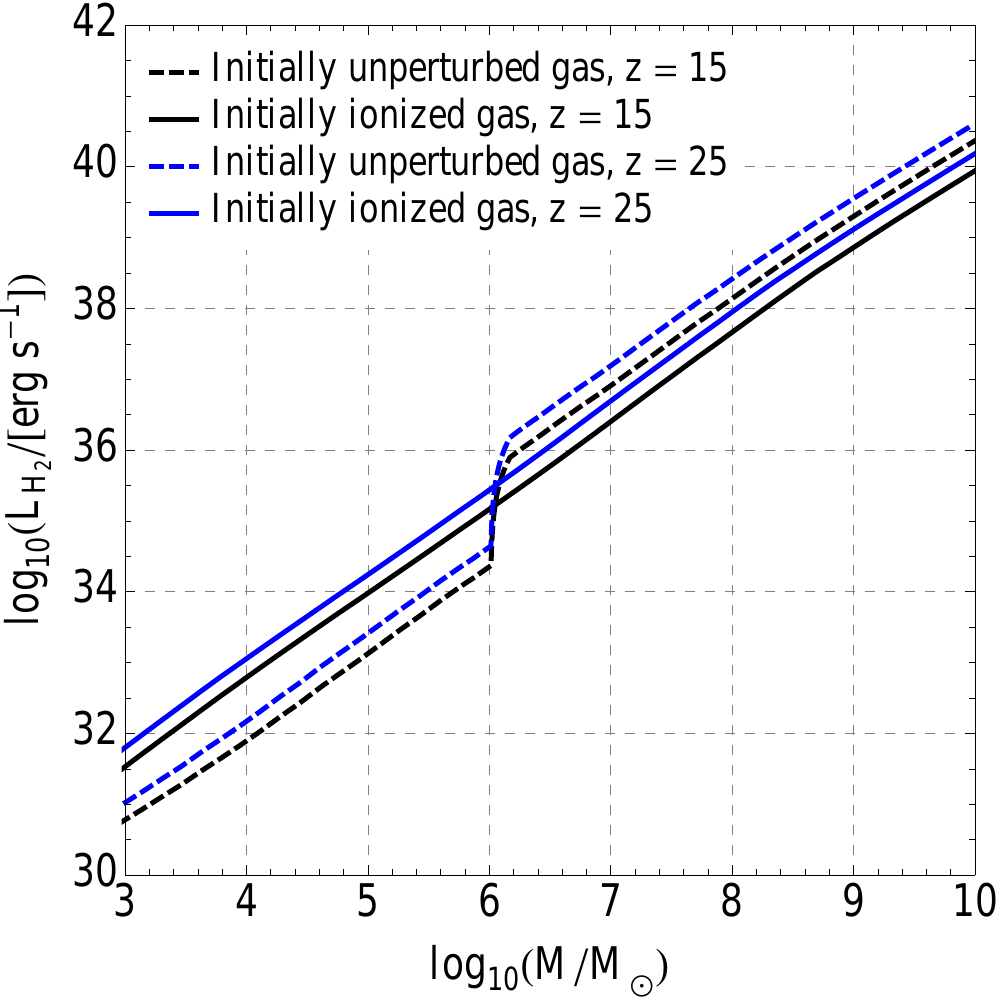}
\caption{\label{fig:LmzH2}
Relationship between \ce{H2}(5-3)  luminosity and halo mass in our model for cosmic dawn, for initially unperturbed ({\em dashed}) and initially ionized ({\em solid}) gas clouds, and at two representative redshifts. In contrast to the \ce{HD} results shown in Fig.~\ref{fig:Lmz}, the relationship is enhanced for unperturbed halos with $M\gtrsim 10^6\msun$ compared to those of lower mass, because the higher temperatures in the more massive halos enhance the collisional cooling rate of \ce{H2}, and this dominates over the mild increase of \ce{H2} abundance in lower-mass halos.
}    
\end{centering}
\end{figure}

We generally follow the modelling approach from Sec.~\ref{sec:cdmodelling}, again taking the initially unperturbed and initially ionized cases as pessimistic and optimistic scenarios for the signal strength. 
We consider collisions of \ce{H2} with \ce{H}, \ce{He}, and \ce{H2}, using the collisional coefficients from Ref.~\cite{LeBourlot1999}\footnote{Available at \url{http://ccp7.dur.ac.uk/cooling_by_h2/}.}. Fig.~\ref{fig:LmzH2} shows the resulting $L_{\rm{H2}}(M,z)$ for each case at $z=15$ and $25$. The \ce{H2} halo luminosity relation is enhanced for higher-mass ($M\gtrsim 10^6\msun$) unperturbed halos, because of the higher temperatures and therefore higher cooling coefficients (see Eqs.~\ref{eq:LambdaXY}-\ref{eq:Lambdan0}) in these halos compared to lower-mass unperturbed halos, which is a stronger effect than the mildly increased \ce{H2} abundances at lower masses (see Fig.~\ref{fig:profiles}). On the other hand, for the \ce{HD} halo luminosities, the higher \ce{HD} abundance at lower halo mass is the dominant effect, resulting in an enhancement of $L_{\ce{HD}}$ at these lower masses (see Fig.~\ref{fig:Lmz}).

We consider an instrument designed to observe at the relevant wavelengths: the Origins Survey Spectrometer (OSS), which will cover 25 to $588\,\mu{\rm m}$ and is planned for the Origins Space Telescope~\cite{Cooray:2019ivm}.  We use OSS specifications from Table 3-1 in Ref.~\cite{Cooray:2019ivm}\footnote{Table 3-1 in Ref.~\cite{Cooray:2019ivm} contains typos in the NEI values for bands~3 and 5, as well as a typo in the stated units~\cite{Bradford-comm}. Our values in Table~\ref{tab:h2inst} correspond to the corrected quantities.}, summarized in our Table~\ref{tab:h2inst}. We consider a 
$0.5\,{\rm deg}^2$ survey lasting 1000 hours, based on the deep extragalactic survey planned for OSS.

\begin{table*}[t]
\centering
\begin{tabular*}{0.85\textwidth}{l c c c c c c c}
\hline
Band & 
Band & 
Central redshift & 
NEI &
Beam &
Number of &
Spectral  &
$P_{\rm N}$ \\
& 
 edges & 
for \ce{H2}(5-3) & 
 &
  FWHM &
spatial pixels &
resolution&
 \\
& 
[$\mu{\rm m}$] & 
 & 
[${\rm MJy}\,{\rm sr}^{-1}\,{\rm s}^{1/2}$] &
[arcsec] &
 &
[GHz] &
[${\rm Jy}^2\,{\rm sr}^{-2}\,h^{3}\,{\rm Mpc}^3$] \\
\hline
\hline\\[-0.5em]
Band 3 & [71, 124] & 9.1 & $1.3$ & 4.0 & 60 & 11  & $4.1\times 10^{8}$ \\
Band 4 & [119, 208] & 16 & $0.85$ & 6.8 & 60 & 6.4  & $1.7\times 10^{8}$ \\
Band 5 & [200, 350] & 27 & $0.44$ & 11 & 48 & 3.8  & $5.1\times 10^{7}$ \\[+0.5em]
\hline
\end{tabular*}
\caption{\label{tab:h2inst}
Specifications of the Origins Survey Spectrometer (OSS) assumed in our forecasts for \ce{H2} intensity mapping at cosmic dawn. We consider a 1000-hour survey of $0.5\,{\rm deg}^2$.
}
\end{table*}

We convert the above information into a 3d noise power spectrum $P_{\rm N}$ using~\cite{Chung:2018szp}
\beq
\label{eq:pN}
P_{\rm N} = \frac{\sigma_{\rm pix}^2}{t_{\rm pix}} V_{\rm vox}\ ,
\eeq
where $\sigma_{\rm pix}$ is the noise-equivalent intensity (NEI). The observing time per spatial pixel, $t_{\rm pix}$, is given by
\beq
\label{eq:tpix}
t_{\rm pix} = \frac{n_{\rm pix} t_{\rm surv}}{\Omega_{\rm surv} / \Omega_{\rm pix}}\ ,
\eeq
where $n_{\rm pix}$ is the number of spatial pixels, $t_{\rm surv}$ and $\Omega_{\rm surv}$ are the observing time and sky area for the entire survey, and $\Omega_{\rm pix}$ is the sky area per pixel. The comoving volume per observed voxel, $V_{\rm vox}$, is given by~\cite{Gong:2011mf}
\beq
V_{\rm vox} = \chi(z)^2 y(z) \Omega_{\rm pix} \delta\nu\ ,
\label{eq:Vvox}
\eeq
where $y(z) \equiv \lambda(1+z)^2 H(z)^{-1}$ converts from (rest) wavelength to radial distance and $\delta\nu$ is the frequency resolution. 

\begin{figure}
\centering
\includegraphics[width=\columnwidth]{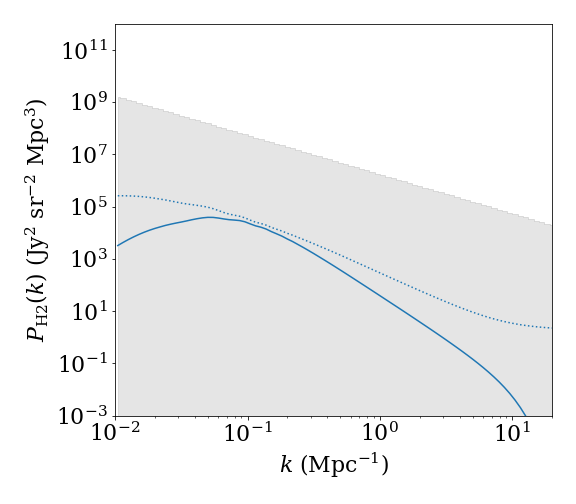}
\caption{$H_2$ power spectrum at $z\sim16$ observed by OSS.  Solid and dashed curves show the signal with and without instrument effects, the shaded region shows the OSS sensitivity.  The all-scales SNR for this observation is $\sim2\times10^{-4}$. 
}
\label{fig:H2_Pk}
\end{figure}

Though Ref.~\cite{Gong:2012iz} found that reasonably high-significance detections of the H$_2$ power spectrum might be possible with an instrument that is nominally similar to OSS, we find that the most up-to-date survey designs for OSS are not as sensitive as the example they used; specifically, we unfortunately find negligible SNRs for all of the  observations we consider.  For example, for an OSS map at 164 $\mu$m (corresponding to $z\sim16$), we find a SNR of $\sim2\times10^{-4}$ (see Fig.~\ref{fig:H2_Pk}).  Though this observation would overlap in redshift with the cosmic dawn HD measurements shown in Fig.~\ref{fig:cosmic_dawn}, the fact that SNRs are low for both HD and \ce{H2} means that even a cross-correlation between the two would not be able to isolate a signal.  Thus, though \ce{H2} remains one of the only other pre-reionization signals available, it would still require futuristic surveys to detect.

\section{Details of cosmic dawn forecasts}
\label{app:cdforecasts}

In this appendix, we collect some details of how our forecasts for cosmic dawn were carried out.
 
\subsection{CMB experiments}

For Planck, we used specifications from Table 4 of Ref.~\cite{Akrami:2018vks}. For PICO, we used the CBE specifications from Table 1.2 of Ref.~\cite{Hanany:2019lle}, while for LiteBIRD, we used Table 1 of Ref.~\cite{HazumiLitebird}; for both of these experiments, we divided the quoted map noise levels by $\sqrt{2}$ to convert from polarization to intensity. For PIXIE, we used Table 1 of Ref.~\cite{Kogut2011}, while our ``enhanced PIXIE" configuration is based on Ref.~\cite{Abitbol:2019ewx}.

For CMB-HD, we used Table 1 of Ref.~\cite{Sehgal:2020yja}. We assumed the 90$\,$GHz and 150$\,$GHz bands to have widths of 27$\,$GHz (the same as the ACT 150$\,$GHz band\footnote{\url{https://act.princeton.edu/technology/specifications}}), and the higher bands to have the same widths as those in FYST, taken from Figure 2 of Ref.~\cite{ChoiCCATp}.

Our noise power spectra for Planck, PICO, LiteBIRD, and CMB-HD can be obtained from the public \texttt{CMBnoise} code\footnote{\url{http://github.com/sjforeman/CMBnoise}}.

\subsection{ALMA}

In our forecasts for ALMA, we used configurations designed to match the characteristics of OSS, but mapped onto observations of \ce{HD} instead of \ce{H2}, as closely as possible, to optimize for cross-correlations of the two tracers. For each frequency band of OSS, we mapped it onto the ALMA band(s) that could observe \ce{HD} at the corresponding redshifts. For each such band, we then computed the sensitivity using the online ALMA sensitivity calculator\footnote{\url{https://almascience.nrao.edu/proposing/sensitivity-calculator}}, with the following settings:
\begin{itemize}
\item {\em observing frequency}: central frequency of ALMA band
\item {\em bandwidth per polarization}: width of ALMA band divided by number of OST spectral pixels
\item {\em number of antennas}: 50 12m antennas (we do not use the 7m antennas because they result in a much larger noise power spectrum for intensity mapping)
\item {\em resolution}: matched as closely as possible to OST beam FWHM values at equivalent frequency, within constraints of synthesized beams achievable by ALMA at given frequency
\item {\em integration time}: 1h (the resulting noise power spectrum can be scaled for other single-observation integration times by multiplying by $[1\,{\rm h}/t_{\rm obs}]$)
\end{itemize}
We convert the resulting sensitivity from $\mu{\rm Jy}\,{\rm beam}^{-1}$ to ${\rm Jy}\,{\rm sr}^{-1}\,{\rm s}^{1/2}$ using the angular resolution and assumed (1h) integration time per observation. We assume that a survey area greater than the field of view will be covered by mosaicked observations spaced by half the primary beam width. The information for each ALMA band, including the derived $P_{\rm N}$, is summarized in Table~\ref{tab:almaspecs}.

\begin{table*}[t]
\centering
\begin{tabular*}{\textwidth}{l c c c c c c c}
\hline
ALMA band & 
Band edges & 
Overlapping &
Spectral &
Synthesized &
Primary&
Sensitivity &
$P_{\rm N}$ \\
& 
& 
mid-IR band & 
resolution &
beam width &
beam width &
for 1\,h observation &
\\
& 
[GHz] & 
& 
[GHz] &
[arcsec] &
[arcsec] &
[$\mu{\rm Jy}\,{\rm beam}^{-1}$]  &
[${\rm Jy}^2\,{\rm sr}^{-2}\,h^{3}\,{\rm Mpc}^3$] \\
\hline
\hline\\[-0.5em]
3 & [84, 116] & OSS band 5 & $0.34$ & 5.1 & 62 & 44 & $6.4\times 10^{2}$ \\
4 & [125, 163] & OSS band 4 & $0.56$ & 3.5 & 43 & 36 & $1.1\times 10^3$  \\
5 & [158, 211] & OSS band 4 & $0.56$ & 2.7 & 33 & 53 & $3.3\times 10^3$  \\
6 & [211, 275] & OSS band 3 & $0.94$ & 2.1 & 25 & 39 & $3.9\times 10^3$  \\
7 & [275, 373] & OSS band 3 & $0.94$ & 1.5 & 19 & 21 & $1.7\times 10^3$  \\[+0.5em]
\hline
\end{tabular*}
\caption{\label{tab:almaspecs}
ALMA configurations and sensitivities used in our forecasts for \ce{HD} intensity mapping at cosmic dawn. We select configurations designed to match the angular resolution and number of frequency channels corresponding to each band of OSS considered in Table~\ref{tab:h2inst}, in order to optimize for \ce{H2}-\ce{HD} cross-correlations. We use a fiducial time of 1 hour per single ALMA observation, and describe in the main text how to estimate the total observing time for a given intensity mapping survey.
}
\end{table*}

ALMA can observe a spectral window up to $7.5\,{\rm GHz}$ wide in a single observation. Thus, if time $t_{\rm obs}$ is devoted to a single observation, the total time to observe a sky area $\Omega$ over a spectral window of width $\Delta\nu$ is given by
\beq
t_{\rm total} = \lp \frac{\Omega}{[0.5\Delta\theta_{\rm pr}]^2} \rp \lp \frac{\Delta\nu}{7.5\,{\rm GHz}} \rp
	t_{\rm obs}\ ,
\eeq
where $\Delta\theta_{\rm pr}$ is the primary beam width from Table~\ref{tab:almaspecs}. The noise power spectrum corresponding to such a survey is the $P_{\rm N}$ value from Table~\ref{tab:almaspecs} multiplied by $(1\,{\rm h}/t_{\rm obs})$.


\bibliography{HD}

\end{document}